\documentclass[aoas,MSNbibl,nameyear,dvips]{arximspdf}
\usepackage{graphicx}

\doi{10.1214/12-AOAS579} %kopijuoti is PTS
\volume{6}
\issue{4}
\pubyear{2012}
\firstpage{1352}
\lastpage{1376}

\makeatletter

\renewcommand{\vec}[1]{\mathbf{#1}}
\newcommand{\vvec}[1]{\bolds{#1}}

\newcommand{\btheta}{\vvec{\theta}}
\newcommand{\bx}{\vec{x}}
\newcommand{\bt}{\vec{t}}
\newcommand{\bu}{\vec{u}}
\newcommand{\bnhat}{\widehat{\vec{n}}}
\newcommand{\bVhat}{\widehat{\vec{v}}}
\newcommand{\G}{\mathrm{G}}
\renewcommand{\L}{\mathrm{L}}
\newcommand{\C}{\mathrm{C}}
\newcommand{\R}{\mathrm{R}}

\makeatother

\begin{document}
\begin{frontmatter}

\title{Inference for population dynamics in the
Neolithic~period\thanksref{T1}}
\runtitle{Population dynamics in the Neolithic}

\thankstext{T1}{Supported in part by Leverhulme Grant F/00125/AD.}

\begin{aug}
\author[A]{\fnms{Andrew W.} \snm{Baggaley}\ead[label=e1]{a.w.baggaley@ncl.ac.uk}},
\author[A]{\fnms{Richard J.} \snm{Boys}\corref{}\ead[label=e2]{richard.boys@ncl.ac.uk}},
\author[A]{\fnms{Andrew} \snm{Golightly}\ead[label=e3]{andrew.golightly@ncl.ac.uk}},
\author[A]{\fnms{Graeme R.} \snm{Sarson}\ead[label=e4]{graeme.sarson@ncl.ac.uk}}
\and
\author[A]{\fnms{Anvar} \snm{Shukurov}\ead[label=e5]{anvar.shukurov@ncl.ac.uk}}
\runauthor{A. W. Baggaley et al.}
\affiliation{Newcastle University}
\address[A]{School of Mathematics and Statistics\\
Newcastle University\\
Newcastle upon Tyne, NE1 7RU\\
United Kingdom\\
\printead{e1}\\
\hphantom{E-mail: }\printead*{e2}\\
\hphantom{E-mail: }\printead*{e3}\\
\hphantom{E-mail: }\printead*{e4}\\
\hphantom{E-mail: }\printead*{e5}} %adresu isvedimo komanda gale!
\end{aug}

% HISTORY:
\received{\smonth{10} \syear{2011}}
\revised{\smonth{6} \syear{2012}}

% ABSTRACT
%
\begin{abstract}
We consider parameter estimation for the spread of the Neolithic
incipient farming across Europe using radiocarbon dates. We model
the arrival time of farming at radiocarbon-dated, early Neolithic
sites by a numerical solution to an advancing wavefront. We allow
for (technical) uncertainty in the radiocarbon data, lack-of-fit of
the deterministic model and use a Gaussian process to smooth spatial
deviations from the model. Inference for the parameters in the
wavefront model is complicated by the computational cost required to
produce a single numerical solution. We therefore employ Gaussian
process emulators for the arrival time of the advancing wavefront at
each radiocarbon-dated site. We validate our model using predictive
simulations.
\end{abstract}

% KEYWORDS
%

\end{frontmatter}

%s1 #&#
\section{Introduction}
\label{secintroduction}

\begin{figure}

\includegraphics{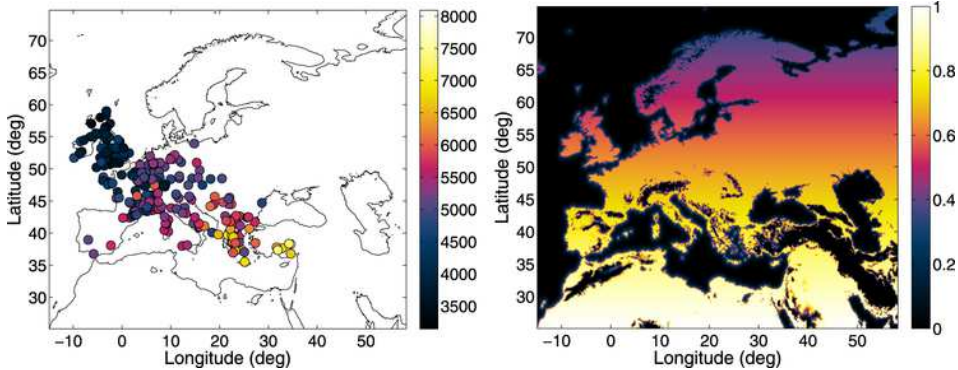}

\caption{Left-hand panel: the location of $^{14}$C-dated early
Neolithic sites used in our analysis and their age (calibrated BC,
%%%ngrs: extraneous reference removed, following comment from Ref. R1.
%colour coded) from $\eqref{eqtisigmai}$. Right-hand panel: The
color coded). Right-hand panel: the
dimensionless diffusivity $\nu_\L$ in the wavefront model
(\protect\ref{nuL}), with a mild linear gradient toward south, local
topographic variations, and the gradual decline to zero away from
the coastlines.}
\label{figrcarb}
\end{figure}

The Neolithic (10,000--4500 BC), the last period of the Stone Age,
was the dawn of a new era in the development of the human race. Its
defining innovation, the transition from foraging to food production
based on domesticated cereals and animals, was one of the most
important steps made by humanity in moving toward modern societies.
Among dramatic changes resulting from the advent of farming were
sedentary living, rapid population growth, gradual emergence of urban
conglomerates, division of labor and the development of complex
social structures. Another important innovation of the period was
pottery making. The mechanism of the spread of the Neolithic remains
an important and fascinating question. The most striking large-scale
feature of this process is its regular character, whereby farming
technologies spread across Europe at a well-defined average speed of
about $U=1$ km$/$year. Impressive and convincing evidence for this
emerged as soon as radiocarbon ($^{14}$C) dates for early Neolithic
sites became available [\citet{Ammerman1971}], and later studies have
confirmed the early results [\citet{Gkiasta2003}].
Figure~\ref{figrcarb} displays the $^{14}$C dates used in our
analysis, and they clearly show the gradual, regular spread of the
Neolithic over a time span of over 4000 years from its origin in the
Near East around 9000 years ago, to the north-west of Europe.

%
%f1 #&#

It is worth emphasising that ``the Neolithic'' is not a single
phenomenon and, although it is often characterized by archaeologists
as a ``package'' of related traits, the individual elements of the
package need not have been transmitted simultaneously. This might
suggest the need to model each distinct element separately. However,
several elements do appear to have traveled together during the
spread of farming into most of Europe [\citet{BurgerThomas2011}] and
the local transition to the full package may have been more rapid than
has often been assumed [\citet{Rowley-Conwy2011}]. Therefore, we
choose to model the Neolithic spread as a single phenomenon.

Despite the overall regular character of the expansion, there are
notable regional variations in the spread of the Neolithic
[\citet{Gkiasta2003,BNLK09}]. Careful analysis of the radiometric and
archaeological evidence has identified various major perturbations:
First, slowing down at topographic obstacles such as mountain ranges.
Second, latitudinal retardation above 54$^\circ$N latitude
[\citet{Ammerman1971}], presumably due to competition with the
pre-existing Mesolithic population, whose density was especially high
in the north [\citet{Fort2010}]; the negative influence of the soil
type and harsher climate in the north also cannot be excluded. Third,
significant acceleration along the Danube--Rhine corridor and the west
Mediterranean [\citet{Ammerman1971}]; see also \citet{Davison2006}.
These variations in the rate of spread complicate quantitative
interpretations of the $^{14}$C data.

Moreover, the data themselves are far from being complete and
accurate. There are inherent errors from the radiocarbon measurement
in the laboratory [\citet{SCN07}] and uncertainties in the conversion
of the $^{14}$C into the calendar age, known as the calibration of the
$^{14}$C dates [\citet{BKLS91,BB08}]. Further errors result from
archaeological factors, such as uncertain attribution of the dated
object to the early farming activity, disturbed stratigraphy of
archaeological sites, etc. [\citet{A90}]. Perhaps even more
importantly, objects related to the first appearance of farming at a
given location are not necessarily related to the first arrival of
farming to its larger region. The occupation of many Neolithic sites
could be a secondary process behind the wave of advance. The only
obvious method to identify such sites is by continuity with its
neighbors.

Given the obviously random nature of the demographic processes
underlying the Neolithic expansion, and the abundance of both random
and systematic errors in the quantitative (mainly radiometric)
evidence available, it is clear that serious statistical analysis is
required to quantify the spread of the Neolithic. All previous work in
this direction has determined the parameter selection for various
models and assessed goodness of fit by using parameter scans. To the
best of our knowledge, the present paper is the first to use formal
statistical inference methods to determine appropriate ranges for
model parameters and assess the fit of the model.

We use a carefully selected set of $^{14}$C dates for the earliest
Neolithic sites across Europe to fit a model of the Neolithic
expansion. The model relies on the fundamental fact, established
empirically as briefly explained above, that the incipient farming
spread at a nearly constant average speed. We also include regional
variations in the speed associated with the latitudinal gradient,
topography and major waterways. Our statistical model adopts an error
structure which accounts for both measurement error in the radiocarbon
dating process and a spatial error process which smooths departures
from the wavefront model. We describe how to obtain the Bayesian
posterior distribution for the unknown model parameters using
simulations from the deterministic model. Simulations of the expanding
Neolithic front are computationally expensive, even at a prescribed
deterministic velocity, precluding their use for Bayesian inference.
Therefore, we construct a Gaussian process emulator
[\citet{SantnerWN03}] for the model, that is, a stochastic approximation
to the arrival times obtained from this model. Such methods are widely
used in the computer models literature; see, for example,
\citet{kennedy01} and references therein. Rather than build a complex
space--time emulator to describe the wavefront model, we pragmatically
build individual emulators for each site and rely on the spatial
Gaussian error process to smooth across geographical space. Finally,
we fit this emulator model to our data and describe the inferences we
draw.

The remainder of this paper is organized as follows. We describe the
$^{14}$C data used in our analysis in Section~\ref{secradioData}. The
implementation of the wavefront model and the statistical model which
links the wavefront model to the data are presented in
Section~\ref{RDWF}. The Gaussian process emulator model and its use
in constructing an inference scheme are the subject of
Sections~\ref{secemulator} and~\ref{secinference}. Finally, the
results of fitting our model to the radiocarbon data are reported in
Section~\ref{secresults}, and we conclude with a discussion in
Section~\ref{secdiscuss}.

%s2 #&#
\section{Selection and analysis of radiocarbon dates}
\label{secradioData}

It is clear from the discussion in Section~\ref{secintroduction} that
$^{14}$C data need to be carefully selected for model parameter
inference. Databases of radiocarbon dates contain several thousand
entries for Neolithic sites across Europe
[e.g., \citet{Gkiasta2003}], but a relative minority of them are
suitable for the studies of the initial spread of the Neolithic. In
this paper, we use carefully selected $^{14}$C dates from 302 earliest
Neolithic sites in Southern and Western Europe from the compilation of
\citet{Davison2009}, where a detailed discussion of the selection
criteria can be found. The objects dated include grains and seeds,
pottery, bone, shells and animal horns, wood, charcoal and peat, and
soil. We note that, in general, there may be issues in using such a
heterogeneous data set, such as those relating to ``chronometric
hygiene'' and the problems that may arise with certain materials or
treatments; see, for example, \citet{Pettitt2003} and the discussion
in \citet{St10}. However, we are not aware of any issues with the
provenance of the samples we use in our analysis; see
\citet{Davison2009} for further details.

The putative origin of the expansion is in an extended region
in the Near East, and, following the results of \citet{Ammerman1971},
later confirmed by many authors, it is reasonable to place it near
Jericho. Similarly to \citet{Davison2009}, the starting date and
position are adopted as 6572 BC and $(\phi,\lambda)=(41.1^\circ$N,
$37.1^\circ$E), the Tell Kashkashok site near Jericho.

Many sites in the selection have several dates (often 3--10, and
occasionally \mbox{30--50}) which can be treated as the arrival time
contaminated by errors. Each date is published together with the
accuracy of radiocarbon measurement in the laboratory, but we recall
that other sources of random and systematic errors are present as
well. Suppose that, at site $i$, we have $m_i$ calibrated $^{14}$C
dates $t_{ij}$ ($j=1,\ldots,m_i$, $i=1,\ldots,n$) and their standard
deviations $\sigma_{ij}$. These data are calculated from the central
99.7\% interval of the calibrated probability distributions for each
object, obtained using calibration curve IntCal04: \citet{intcal04}.
Specifically, each $t_{ij}$ is taken to be the midpoint of its
calibrated interval and $\sigma_{ij}$ to be one-sixth of the length of
the interval. This data set can be found in the supplementary material
[\citet{Baggaley2013}]
for this paper. We can calculate summaries for the data at each site,
namely, an arrival time $t_i$ and its accuracy $\sigma_i$, as defined
in this context by \citet{SCN07}, by appropriately weighting the data,
as
%
%e1 #&#
\begin{equation}
\label{eqtisigmai} t_i= \sum_{j=1}^{m_i}t_{ij}
\sigma^{-2}_{ij}\Big/\sum_{j=1}^{m_i}
\sigma^{-2}_{ij} \quad\mbox{and}\quad \sigma^2_i=1\Big/
\sum_{j=1}^{m_i} \sigma^{-2}_{ij}.
\end{equation}

Clearly, there is some loss of information in not using the full
calibrated probability distribution for each object. However, to
include this within our analysis by using, for example, a normal
mixture model representation for each distribution, would make our
analysis considerably more complicated. In any case, the model we do
use in our analysis is one for the weighted means $t_i$, whose
calibrated distributions will be roughly normally distributed due to
the central limit theorem. The left-hand panel of
Figure~\ref{figrcarb} shows the sites on the map, color-coded
according to the arrival time $t_i$.

%s3 #&#
\section{Justification and implementation of the wavefront model}
\label{RDWF}
The wave of advance model\setcounter{footnote}{1}\footnote{To avoid confusion, we note that
the word ``wave'' does not refer here to any oscillatory behavior.}
has a deep mathematical basis. Such fronts are the salient feature of
a wide class of phenomena that involve an exponential growth of some
quantity (in this case, the population density) and its spread via
random walk or related diffusion; see \citet{FP08} and references
therein. This behavior is captured by the classical equation of
\citet{F37} and \citet{Kolmogorov1937}, known as the FKPP equation,
of the form
%
%e2 #&#
\begin{equation}
\label{eqFKPP} \frac{\partial N}{\partial t} =\gamma N \biggl(1-\frac{N}{K} \biggr) +
\nabla\cdot(\nu_\G\nabla N),
\end{equation}
where $N$ is the population density, a function of position $\bx$ and
time $t$, $\gamma$ is the growth rate of the population (the
difference of the birth and death rates per unit area), $K$ is the
carrying capacity of the landscape (the maximum sustainable population
density), and $\nu_\G$ is the diffusivity, a measure of human
mobility. For constant~$K$, this equation has obvious solutions $N=0$
and $N=K$. Solutions of the initial value problem for this equation in
a homogeneous domain, with a localized initial condition, have the
form of a wave of advance, $N(x-Ut)$ in one dimension, where the
solution changes from $N=0$ ahead of the wavefront to $N=K$ behind it.
The width of the transition region (an internal boundary layer) is of
order $d=\sqrt{\nu_\G/\gamma}$ and, in one dimension, the front
propagates at a constant speed
%
%e3 #&#
\begin{equation}
\label{eqfrontspeed}
U=2\sqrt{\gamma\nu_\G}
\end{equation}
independently of $K$. In two dimensions, the constant-speed
propagation provides a good approximation as soon as the radius of
curvature of the front becomes much larger than $d$. For parameter
values characteristic of the Neolithic expansion---for example,
$\gamma\simeq0.02$ year$^{-1}$ and $\nu_\G\simeq15$
km$^2$ s$^{-1}$ (see the discussions in Sections~\ref{sectregvar}
and~\ref{sectbayesinf}), providing $U\simeq1$ km s$^{-1}$---the
front width is $d\simeq30$ km, and the model of a wavefront
propagating at a constant speed is safely applicable at the
continental scales of 100--1000 km. Even though we aim at an
inference for the speed of the Neolithic expansion, we find it
convenient to present our results with reference to the FKPP model. In
particular, $U$ is parameterized in terms of $\gamma$ and $\nu_\G$
since these quantities admit a direct interpretation in terms of human
behavior. Nevertheless, we emphasize that the inference that follows
does not depend in any fundamental way on the FKPP equation, or on the
assumptions of the demographic processes underlying it; the results we
present may simply be considered as providing statistical constraints
on the speed of a wavefront $U$, independently of any interpretation
of the mechanisms behind this wave.

%s3.1 #&#
\subsection{The wavefront representation}
A discrete version of the wavefront model, suitable for numerical
simulations, is constructed as follows. We select the point source of
the population and then, at a small radius from this point, we arrange
a small number of test points (``particles'') on a circle around it,
with the position of particle $i$ denoted by
$\bx_i=(\phi_i,\lambda_i)$, where $\phi_i$ and $\lambda_i$ are the
geographical latitude and longitude, respectively. This allows us to
compute the local tangent and normal vectors of the front, in
particular, the local unit outward normal $\bnhat_i$ (in the direction
of the advancing front). At each time step we move each particle
along $\bnhat_i$ with the \textit{local} velocity $\bu$ given by
(\ref{bu}). The wavefront thus expands outward in time; in a
homogeneous, isotropic domain, the expansion would take the form of
concentric circles, but in the heterogeneous and anisotropic domain
used here (described in detail below), the expansion is irregular and
tracks the front obtained from the corresponding FKPP model. At the
continental scale, the Earth's curvature is noticeable and we work on
a spherical surface of the appropriate radius.

The separation of the particles increases as the front expands, and an
additional particle is introduced between any two particles if their
separation exceeds a specified value $\delta$. Conversely, if the
front contracts, particles that are closer than $\delta/2$ are
removed. We take $\delta$ to be equal to the grid spacing (introduced
below) of 4~arc-minutes, that is, between 3 km at 60$^\circ$N and
7 km at 30$^\circ$N. It is important to monitor the particle ordering
along the front after adding or removing particles in order to
maintain a continuous wavefront.

When the front encounters an obstacle (e.g., a mountain range), its
flanks move round it and then have to merge behind the obstacle. The
merger is implemented by applying algorithms initially used to model
the evolution of magnetic flux tubes in astrophysical simulations, and
so described in more detail in \citet{Baggaley2009}. Briefly, we
check, at each time-step, if any two particles (which are not
neighbors) are closer than $\delta$; if so, the particle ordering is
switched to ``short-circuit'' any (almost closed) loops that have
arisen. The front then propagates further, leaving behind closed loops
which are then removed from the simulation since they do not affect
the front arrival time.

%s3.2 #&#
\subsection{Regional variations}\label{sectregvar}

The front speed (\ref{eqfrontspeed}) depends on the product of
$\gamma$ and $\nu_\G$, so that these parameters cannot be estimated
simultaneously from studies of the front propagation. The growth rate
is rather well constrained by biological and anthropological factors:
\citet{Birdsell1957} reports values in the range 0.029--0.035
year$^{-1}$\vspace*{1pt} and \citet{Steele1998} suggest the range
0.003--0.03 year$^{-1}$. Here, for comparability with our
own recent studies [e.g., \citet{Davison2006}], we take
$\gamma\simeq0.02$ year$^{-1}$, corresponding to a generation
time of 30 years. This choice does not limit the applicability of our
model. If, for example, a slightly different value of $\gamma$ is
preferred, then the reported values of $\nu_\G$ need only to be scaled
so that the product $\gamma\nu_\G$ remains unchanged.

To make the problem tractable, we adopt a specific spatial variation
of the diffusivity and seek inference for its magnitude, just a single
scalar parameter. Such assumptions unavoidably involve significant
arbitrariness; however, a well-informed assumption is justifiable if
it results in a model that fits the observations to a satisfactory
degree. Thus, we represent the diffusivity in the form
%
%e4 #&#
\begin{equation}
\label{nuG} \nu_\G(\phi,\lambda)=\nu \nu_\L(\phi,
\lambda),
\end{equation}
where $\nu$ is the constant-dimensional magnitude and $\nu_\L$ is a
dimensionless function of position whose magnitude is of order unity
and whose form is assumed to be known. We prescribe $\nu_\L$ to depend
solely on the topography (altitude above the sea level) and latitude
as described below.

Early farming does not appear to have been practical at high altitudes
in the temperate zones considered here [e.g., sites in the
Alpine foreland are not found on land above 1 km:
\citet{Whittle1996}];
to include this effect in our model, we smoothly reduce $\nu_\L$ to
zero at around this height. Also, in order to allow for coastal sea
travel, $\nu_\L$ exponentially decreases with the distance $d_{\mathrm L}$
to the nearest land. Finally, we include a linear decrease of $\nu_\G$
toward the northern latitudes. Altogether, we adopt the following
form for the dimensionless diffusivity:
%
%e5 #&#
\begin{equation}
\label{nuL} \nu_\L= \biggl(1.25-\frac{\phi}{100^\circ} \biggr)\times
\cases{\frac{1}{2}-\frac{1}{2}\tanh\bigl\{10(a-1 \mbox{
km})\bigr\},&\quad
$a>0$,
\vspace*{2pt}\cr
\exp(-d_{\mathrm L}/10 \mbox{ km}), &\quad $a<0$,}
\end{equation}
where $a$ is the altitude, $\phi$ is the latitude, and we recall that
$a<0$ corresponds to the seas. Both $a$ and $d_{\mathrm L}$ are
complicated functions of position reflecting the topography and
coastlines of Europe. Of course, the choice of these particular forms
are only a crude and exploratory attempt to capture the dependence on
latitude and altitude. However, we have found our results not to be
too sensitive to minor changes in these functions.

The altitude data have been taken from the ETOPO1 1-minute Global
Relief database [\citet{geodas}]. As a reasonable compromise between
computational efficiency and spatial accuracy, we use a
$740\times1100$ grid with a spatial resolution of 4 arc-minutes
between $15^\circ$W and $60^\circ$E in longitude and $25^\circ$N
and $75^\circ$N in latitude. The boundaries of the domain explored
and the spatial variation of $\nu_\L$ are shown in
Figure~\ref{figrcarb}.

Significant regional anomalies revealed by archaeological and
radiometric evidence are the early Neolithic Linearbandkeramik (LBK,
Linear Pottery) Culture that propagated in 5500--4900 BC along the
Danube and Rhine rivers at a speed of at least 4 km$/$year
[\citet{DSGSTZ05}] and the Impressed Ware culture (5000--3500 BC) that
spread along the west Mediterranean coastline at an even higher speed.
The latter could be as high as 10 km$/$year [\citet{Zilhao2001}].
\citet{Davison2006} included directed spread (advection) along the
major river paths and coastlines into their mathematical model to
achieve a significant improvement in the fit to $^{14}$C dates.

We include advection along the Danube--Rhine river system and the
coastlines using world map data from \citet{CIA}, in the form of
tangent vectors of the coastlines and river paths, $\bVhat_\C$ and
$\bVhat_\R$, respectively, given at irregularly spaced positions. To
remap the data to the regular grid introduced above, we use the
original data weighted with $\exp(-d_V/15\mbox{ km})$, where $d_V$ is
the distance between the grid point and the data point, and the chosen
scale of 15 km contains 2--5 grid separations at latitudes
30$^\circ$--60$^\circ$N.

With the unit tangent vectors for the coastlines and river paths thus
defined, the magnitudes of the respective local front velocities,
$V_\C$ and $V_\R$, are subject to inference. In other words, the local
front velocity is given by
%
%e6 #&#
\begin{equation}
\label{bu} \bu=U \bnhat+ \vec{V},
\end{equation}
where
$\vec{V}=V_\C \operatorname{sgn}(\bnhat\cdot\bVhat_\C) \bVhat_\C
+ V_\R \operatorname{sgn}(\bnhat\cdot\bVhat_\R) \bVhat_\R$.
Since, unlike the geographical data, the points defining the front are
not necessarily at the nodes of the regular grid, we used bilinear
interpolation from the four closest grid points to calculate $\nu_\L$
(and thus $U$), $\bVhat_\C$ and $\bVhat_\R$ at the front positions. We
note that the accuracy of the wavefront model was tested by comparing
it with numerical solutions of (\ref{eqFKPP}) for a wide range of
parameter values; the agreement was invariably quite satisfactory with
a typical deviation of 40 years.

%s3.3 #&#
\subsection{Statistical model}

We model the radiocarbon dates for each object at each site as being a
noisy version of those dates predicted by the wavefront model,
explicitly allowing for three types of error introduced in
Section~\ref{secintroduction}: the date $t_{ij}$ of object $j$ at a
site $i$ (location $\bx_i$), for $j=1,\ldots,m_i$, $i=1,\ldots,n$, is
modeled as
\[
t_{ij}=\tau_i(\btheta)+\zeta(
\bx_i)+\sigma_{ij} \omega_{ij} +\sigma
\varepsilon_{ij},
\]
where\vspace*{1pt} $\tau_i(\btheta)$ is the deterministic wavefront arrival time, a
function of $\btheta=(\nu,V_\C,V_\R)^T$, and $\omega_{ij}$ and
$\varepsilon_{ij}$ are independent error terms following a standard
normal distribution. Here\vspace*{1pt} $\zeta(\bx_i)$ allows for spatially
inconsistent data, $\sigma_{ij}$ is the standard deviation of the
$^{14}$C measurement in the laboratory and calibration, and $\sigma$
is a global error term. Between them, the $\zeta$ and $\sigma$ terms
allow for mismatches between our model and the data: $\zeta$ allows
for a certain level of variability between the arrival times at nearby
sites, while $\sigma$ allows for an additional global variability
(with no spatial correlation). As well as the mismatch arising when a
simple large-scale model is applied to a process with inherent
variability on smaller scales, these terms will also allow our model
to be fairly robust to any problems in our radiocarbon data such as
the misinterpretation of sites of secondary Neolithic occupation as
being indicative of the wave of first arrival. We note that some
authors [e.g., \citet{BKLS91}] choose to model the start-date of a
``first arrival phase'' using an additional parameter for each site,
with the explicit expectation that all individual samples will fall
after this date. However, our wavefront model attempts to capture the
Neolithic transition over much larger spatial and temporal scales and
so we choose not to incorporate such detailed temporal effects.

We model the spatially smooth error process $\zeta$ using a zero mean
Gaussian process (GP) prior [\citet{Rasmussen2006}] with covariance
function $k_\zeta(\cdot,\cdot)$, that is,
\[
\zeta(\cdot)\sim\operatorname{GP}\bigl(\vec{0},k_{\zeta}(\cdot,\cdot)
\bigr).
\]
We impose smoothness by taking a Gaussian kernel for the covariance
function so that the covariance between spatial errors at locations
$\bx$ and $\bx'$ is
\[
k_{\zeta}\bigl(\bx,\bx'\bigr)=a_\zeta^2
\exp \bigl\{-\bigl(\bx-\bx'\bigr)^T\bigl(\bx-
\bx'\bigr)/r_\zeta^2 \bigr\}.
\]
The parameters of this function control the overall level of
variability and smoothness of the process, with larger values of the
length scale $r_\zeta$ giving smoother realizations and thereby
down-weighting sites with different dates relative to nearby sites.

In the following analysis it will be more straightforward to work with
an equivalent model for the summary statistics $(t_i,\sigma_i)$ in
(\ref{eqtisigmai}), namely,
%
%e7 #&#
\begin{equation}
\label{statmodel} t_i=\tau_i(\btheta)+\zeta(
\bx_i)+\sigma_i\omega_i+\sigma
\varepsilon_i,\qquad i=1,\ldots,n,
\end{equation}
where the $\omega_i$ and $\varepsilon_i$ are independent error terms
following a standard normal distribution. Thus, the inferential task
is to make plausible statements about the wavefront model parameters
$\btheta$, the global error term $\sigma$ and the Gaussian process
hyperparameters $a_\zeta$ and $r_\zeta$.

%s3.4 #&#
\subsection{Bayesian inference}
\label{sectbayesinf}
We adopt a Bayesian approach to inference and express our prior
uncertainty for the unknown quantities through a density
$\pi(\btheta,\sigma,a_\zeta,r_\zeta)$. The speed of spread reported
by \citet{Ammerman1971}, together with the growth rate discussed in
Section~\ref{sectregvar}, suggests $\nu\simeq12$ km$^2$\mbox{$/$}year.
The spread reported for the LBK culture
[e.g., \citet{Ammerman1973,Gkiasta2003}] suggests $V_\R\simeq
5$ km$/$year. The spread reported for the Impressed Ware culture
[\citet{Zilhao2001}] would suggest $V_\C\simeq10$ km$/$year.
However, this study is restricted to the Western Mediterranean and we
believe this parameter would take a smaller value over the European
continent such as $V_\C\simeq3$ km$/$year.
We do not expect the wavefront model to fit the data well in all regions,
so we expect a relatively large value of $\sigma$
(compared with the accuracy of individual dates),
$\sigma\simeq500$ years.
The magnitude of
$a_\zeta$, which models the variability in dates between spatially
nearby sites, is difficult to estimate, but the maximum accuracy of
laboratory radiocarbon measurements (which will probably be the
smallest source of variability within our model) suggests a minimum
significant value of $a_\zeta\simeq20$ years. The mean
minimum separation of radiocarbon-dated sites suggests
$r_\zeta\simeq1.3^\circ$. We use the values above to guide our
choice of mode for the prior distribution. In view of the
uncertainties in these values, we make the prior rather diffuse, with
independent components. Specifically, we use the log-normal and
inverse-gamma distributions, with
\begin{eqnarray*}
\nu&\sim&\operatorname{LN}(3.5,1),\qquad V_\C\sim\operatorname{LN}
\bigl(1,0.5^2\bigr),\qquad V_\R\sim\operatorname{LN}(2.6,1),
\\
\sigma^2&\sim&\operatorname{IG}\bigl(5,10^6\bigr),\qquad
a_\zeta\sim\operatorname{LN}\bigl(5,1.5^2\bigr),\qquad
r_\zeta\sim\operatorname{LN}\bigl(2.5,1.5^2\bigr).
\end{eqnarray*}
Let $\bt=(t_{1},\ldots,t_{n})^T$ and
$\vvec{\tau}(\btheta)=(\tau_{1}(\btheta),\ldots,\tau_{n}(\btheta))^T$
denote the observed arrival times and those from the wavefront model
(evaluated at $\btheta$). Then, using (\ref{statmodel}), the data
model is an $n$-dimensional normal distribution, with
\[
\bt|\btheta,\sigma,a_\zeta,r_\zeta\sim \mathrm{N}_n
\bigl(\vvec{\tau}(\btheta),\Sigma\bigr),
\]
where $\Sigma=K_{\zeta}(X,X)+\operatorname{diag}(\sigma^2_i+\sigma^2)$,
$K_{\zeta}(X,X)$ has $(i,j)$th element $k_{\zeta}(\bx_i,\bx_j)$, and
$X=(\bx_{1},\ldots,\bx_{n})^{T}$ are the locations of the
radiocarbon-dated sites. Combining both sources of information
(the data/model and the prior) gives the joint posterior density as
\[
\pi(\btheta,\sigma,a_\zeta,r_\zeta|\bt)\propto \pi(\btheta,
\sigma,a_\zeta,r_\zeta) \pi(\bt|\btheta,\sigma,a_\zeta,r_\zeta).
\]
In practice, the posterior density is analytically intractable and we
therefore turn to a sampling-based approach to make inferences. Markov
chain Monte Carlo (MCMC) methods can readily be applied to this
problem, but such schemes will typically need many thousands or
millions of evaluations of $\vvec{\tau}(\btheta)$, each requiring a
full simulation of the wavefront model expanding across the whole of
Europe. This takes around 10 seconds on a quad-core CPU with a clock
speed of 2.67 GHz and the code parallelized using OpenMP. This
computational cost prohibits the use of an MCMC scheme for sampling
from the posterior distribution of the unknown quantities. To proceed,
we therefore seek a faster approximation of the first arrival times
from the wavefront model.

%s4 #&#
\section{An approximation to the wavefront model}
\label{secemulator}
The first arrival times from the wavefront model can be approximated
by using both deterministic and stochastic methods such as cubic
splines and Gaussian processes, respectively. These approximations
(known as emulators) are determined by first setting up an
experimental design consisting of a number of locations and parameter
values, and then computing the (deterministic) first arrival times
from the wavefront model using the locations and parameter values in
the design. Finally, the proposed emulator is fitted to this model
output. In line with many other authors, for example, \citet{kennedy01}
and \citet{henderson09}, we favor using Gaussian processes to emulate
the first arrival times, as they not only fit this model output exactly
and interpolate smoothly between design points but also quantify
levels of uncertainty around interpolated values. Rather than build a
complex emulator whose inputs are both site location and wavefront
model parameters, we pragmatically build individual (parameter only)
emulators for each site and rely on a spatial Gaussian error process
to smooth the resulting spatial inhomogeneities. This strategy is
viable because the inference scheme described later in
Section~\ref{secinference} only requires emulation at the 302 sites
in the radiocarbon data set. One possible drawback is that it does
produce site-specific emulators that are not as spatially smooth as
the more complex emulator. However, there are significant
computational benefits of building separate emulators for each of the
302 sites, as they work in a smaller input space and can be computed in
parallel.

%s4.1 #&#
\subsection{A Gaussian process emulator}
\label{GPem}
The output from a single simulation of the wavefront model with input
parameter values $\btheta=(\nu,V_\C,V_\R)^T$ gives the front arrival
time for all sites. Consider the arrival time $\tau_{i}(\btheta)$ at
a single site $i$. We model the emulator for the arrival time at this
site using a Gaussian process with mean function $m_{i}(\cdot)$ and
covariance function $k_{i}(\cdot,\cdot)$, that is,
%
%e8 #&#
\begin{equation}
\label{eqGPrelation} \tau_{i}(\cdot) \sim\operatorname{GP}
\bigl(m_{i}(\cdot),k_{i}(\cdot,\cdot)\bigr).
\end{equation}
A suitable form for the mean function can be determined by noting that
the great-circle distance $d$ from the source of the wavefront is
approximated by $d=u\tau$, where $u$ is the front speed given by
(\ref{bu}) and $\tau$ is the arrival time. Hence, $\tau\propto1/u$
and therefore we seek
\[
m_{i}(\vvec{\theta})= \alpha_{i,0}+\alpha_{i,1}
\frac{1}{\sqrt{\nu}}+\alpha_{i,2}\frac
{1}{V_\C}+\alpha_{i,3}
\frac{1}{V_\R},
\]
where $\alpha_{i,k}$ are coefficients to be determined, which account
for the relative importance of diffusivity, coastal and river speeds.
We determine them using ordinary least squares fits. Note that the mean
function depends on $\sqrt\nu$ to be consistent
with~(\ref{eqfrontspeed}). We use a stationary Gaussian covariance
function
\[
k_{i}\bigl(\vvec{\theta},\vvec{\theta}'\bigr)
=a_i^2\exp \Biggl\{-\sum_{j=1}^3
\bigl(\theta_j-\theta_j'
\bigr)^2/r^{2}_{i,j} \Biggr\},
\]
where $r_{i,j}$ is the length scale at site $i$ along the
$\theta_j$-axis, as is widely used in similar applications since it
leads to realizations that vary smoothly over the input space
[\citet{SantnerWN03}]. We describe how to make inferences about the
parameters $(a_{i},\vec{r}_{i})$ of these site-specific covariance
functions in Section~\ref{secemFit}.

Suppose that $p$ simulations of the (computationally expensive)
wavefront model are available to us, each providing the arrival time
at each radiocarbon-dated site. Let $\vvec{\tau}_{i}(\Theta)
=(\tau_{i}(\btheta_{1}),\ldots,\tau_{i}(\btheta_{p}))^{T}$ denote the
$p$-vector of arrival times resulting from the wavefront model with
input values $\Theta=(\vvec{\theta}_1,\ldots, \vvec{\theta}_{p})^{T}$,
where $\btheta_i=(\nu_i,V_{\C,i},V_{\R,i})^T$. From
(\ref{eqGPrelation}) we have
\[
\vvec{\tau}_{i}(\Theta) \sim N_p\bigl(
\vec{m}_{i}(\Theta),K_{i}({\Theta},{\Theta})\bigr),
\]
where $\vec{m}_{i}(\Theta)$ is the mean vector with $j$th element
$m_{i}(\vvec{\theta}_{j})$ and $K_{i}({\Theta},{\Theta})$ is the
variance matrix with $(j,\ell)$th element
$k_{i}(\vvec{\theta}_{j},\vvec{\theta}_\ell)$.

We can use the simulations to model the front arrival times
corresponding to other values of the input parameters. Indeed, suppose
we need to simulate front arrival times at a collection of $p^*$ new
design points $\Theta^{*}=(\btheta_{1}^*,\ldots,\btheta^*_{p^{*}})^{T}$. This is straightforward as, with the local
emulators, the arrival times are mutually independent and the
distribution of the arrival time $\vvec{\tau}_i(\Theta^*)$ at
site $i$ conditional on the training data
$\vvec{\tau}_{i}(\Theta)$ can be determined using standard
properties of the multivariate normal distribution as
%
%e9 #&#
\begin{equation}
\label{emPred} \vvec{\tau}_i\bigl(\Theta^*\bigr)|\vvec{
\tau}_{i}(\Theta) \sim N_{p^*} \bigl(\vvec{\mu}_{i}
\bigl(\Theta^*\bigr),V_{i}\bigl(\Theta^*\bigr) \bigr),
\end{equation}
where
%
%e10 #&#
%e11 #&#
\begin{eqnarray}
\label{emMean} \vvec{\mu}_{i}\bigl(\Theta^*\bigr) &=&
\vec{m}_{i}\bigl(\Theta^*\bigr)+K_{i}\bigl(\Theta^*,\Theta
\bigr)K_{i}(\Theta,\Theta )^{-1} \bigl\{\vvec{
\tau}_{i}(\Theta) -\vec{m}_{i}(\Theta) \bigr\},
\\
\label{emVar} V_{i}\bigl(\Theta^*\bigr) &=& K_{i}\bigl(
\Theta^*,\Theta^*\bigr)-K_{i}\bigl(\Theta^*,\Theta\bigr)K_{i}(
\Theta, \Theta;a_{i},\vec{r}_{i})^{-1}K_{i}
\bigl(\Theta,\Theta^*\bigr).
\end{eqnarray}
Note that, to simplify the notation, we have dropped the dependence in
these expressions on the hyperparameters $(a_{i},\vec{r}_{i})$,
where $\vec{r}_{i}=(r_{i,1},r_{i,2},r_{i,3})^{T}$.

%s5 #&#
\section{Inference}
\label{secinference}

Before we can fit the Gaussian process emulators, we need to obtain
the training data from the wavefront model. These data are obtained by
running the emulators at a particular experimental design; see
Section~\ref{secexptdesign} for details. Given the $^{14}$C arrival
times and the training data, it is possible in theory to fit the
emulator and statistical model (\ref{statmodel}) jointly using an
MCMC scheme. However, this is likely to be extremely computationally
expensive, and we therefore fit the emulator and the statistical model
separately. This approach is advocated by \citet{bayarri07} and adopted
by \citet{henderson09}, among others. Details on how we fit the
emulators and demonstrate their accuracy can be found in
Sections~\ref{secemFit} and~\ref{secemVal}.

We are now in a position to build an inference algorithm using the
site-specific stochastic emulators $\tau_i^*(\btheta)$ instead of the
deterministic wavefront model $\tau_i(\btheta)$. In doing so, we fix
each emulator's hyperparameters $(a_i,\vec{r}_i)$ at their posterior
mean. It is possible to undertake inferences allowing for uncertainty
in the hyperparameters. We describe methods for doing this in
Section~\ref{sechyperuncer} and also show that allowing for such
uncertainty in our analysis has only a very minor affect on the
posterior distribution and hence on our inferences.

Using the subscript $e$ to denote the use of the emulators, the
(joint) posterior density for the unknown parameters in the
statistical model, replacing (\ref{statmodel}), is
%
%e12 #&#
\begin{equation}
\label{jpStatM} \pi_{e}(\btheta,\sigma,a_\zeta,r_\zeta|
\bt)\propto \pi(\btheta,\sigma,a_\zeta,r_\zeta)
\pi_{e}(\bt|\btheta,\sigma,a_\zeta,r_\zeta),
\end{equation}
where
\[
\pi_{e}(\bt|\btheta,\sigma,a_\zeta,r_\zeta) \propto
\bigl|\Sigma(\btheta)\bigr|^{-1/2} \exp \bigl\{- \tfrac{1}{2} \bigl(\bt-
\vvec{\mu}(\btheta)\bigr)^T\Sigma(\btheta)^{-1} \bigl(\bt-
\vvec{\mu}(\btheta)\bigr) \bigr\},
\]
$\vvec{\mu}(\btheta)$ has $i$th element $\mu_{i}(\btheta)$ and, to
allow for emulator uncertainty, we redefine
$\Sigma(\btheta)=K_{\zeta}(X,X)+\operatorname{diag}\{V_{i}(\btheta
)+\sigma^2_i+\sigma^2\}$,
where $X=(\bx_1,\ldots,\bx_n)^T$. Sampling from the posterior
distribution (\ref{jpStatM}) can be achieved by using a
Metropolis--Hastings scheme similar to that described in
Section~\ref{secemFit}. Specifically, we use correlated normal
random walks to propose updates on a log-scale in separate blocks for
$\btheta$, $\sigma$ and the parameters $(a_\zeta,r_\zeta)$ that govern
spatial smoothness. The covariance structure of the innovations was
chosen using a series of short trial runs.

%s6 #&#
\section{Results}
\label{secresults}
\begin{figure}

\includegraphics{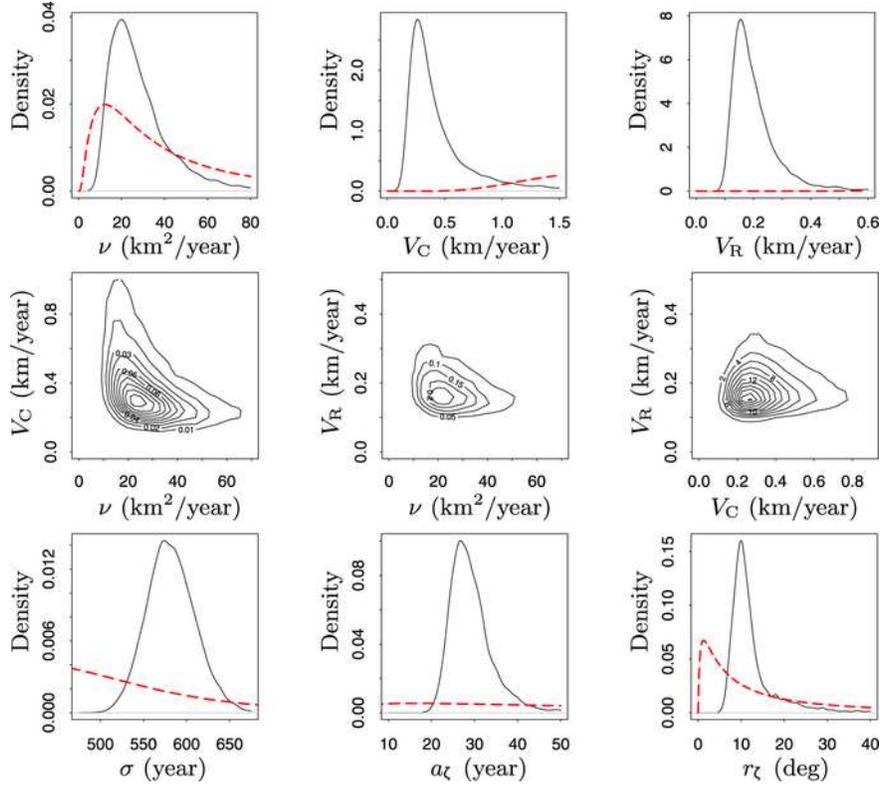}

\caption{Marginal (upper row, from left to right) and joint
(middle row) posterior densities for the wavefront model
parameters $\nu$, $V_\C$ and $V_\R$, and marginal posterior
densities (bottom row, left to right) for the residual
uncertainty parameter $\sigma$ and the spatial GP
hyperparameters $a_\zeta$ and $r_\zeta$. The prior densities are
shown in dashed red.}
\label{figpostplots}
\end{figure}

We found that running the MCMC scheme for 1M iterations after a
burn-in of 100K iterations and thinning by taking every 100th iterate
gave a sample of $J=10$K (effectively uncorrelated) values from the
posterior distribution. Figure~\ref{figpostplots} shows the marginal
and joint posterior densities of the key parameters of the wavefront
model (the diffusivity $\nu$, the coastal velocity $V_\C$ and the
river-path velocity $V_\R$) and also the parameters of the
statistical model (the global error $\sigma$, and the Gaussian Process
hyperparameters $a_\zeta$ and $r_\zeta$). The reduction in
uncertainty from the prior to the posterior, which is considerable for
most parameters, shows that the data have clearly been informative.

%f2 #&#

The posterior distributions of the wavefront model are in reasonable
agreement with the range of values suggested by previous studies. For
example, the mode of the marginal posterior for the diffusivity $\nu$
is around 20 km$^2$\mbox{$/$}year, which corresponds to a global rate of
spread of $U=2\sqrt{\gamma\nu}\simeq1.25$ km$/$year. The modal values
for the amplitudes of the advective velocities, $V_\C=0.3$ km$/$year
and $V_\R=0.2$ km$/$year, are rather lower than the values suggested in
the archaeological literature which motivated their study
[\citet{Zilhao2001,DSGSTZ05}]. Regarding the coastal speed $V_\C$,
this could be explained by an accelerated spread only being required
in the west Mediterranean, whereas our model applies it to the whole
European coastline; an alternative model, allowing regional variations
in coastal effects, may produce different results. The relatively
small value for $V_\R$ is perhaps more surprising, given the
prevailing archaeological picture of relatively rapid spread of the
LBK culture in the Danube and Rhine basins. It may be that the
earliest dates associated with the LBK culture do not correspond
particularly well to spread as a continuous wave and that an
alternative model might also be preferable here. After taking account
of the overall level of fit of our current model, the dates within
this region are satisfactorily explained with a relatively weak
advective enhancement of the background diffusive spread. It is also
plausible that the slightly enhanced value of $\nu$ (compared with
many of the studies cited in Section~\ref{secintroduction}) results
in the relatively low values for $V_\C$ and $V_\R$.

The marginal posterior distribution for the global error
parameter $\sigma$ is centred on a value of around 575 years (see
Figure~\ref{figpostplots}). This value quantifies the mismatch
between our mathematical model of the spread and the local variations
present in the actual spread. Thus, our inference suggests that a
simple large-scale wave of advance, while remaining a good model on
the continental scale, should only be considered a good model on time
scales of order 600 years (and thus length scales of order 600 km) or
greater; on shorter time scales, significant local variations should
be expected. Mismatch between model and data is also allowed in our
spatial term, $\zeta$; the marginal posterior distributions for its
amplitude ($a_{\zeta}$) and length scale ($r_{\zeta}$) are relatively
tightly peaked around values of order 30 years and $10^\circ$,
respectively. The latter value suggests that correlations between
dates at nearby sites are effectively significant within distances of
order 750 km. (Note that this is approximately the length scale on
which $\sigma$ suggests that the variations should be expected within
the large scale model, so that the two results are consistent.)
However, the amplitude $a_{\zeta}$ is relatively small compared
with $\sigma$, suggesting that ``within region'' variation is not
particularly significant, in comparison with larger scale deviations
from the model.

We can examine the operation of the spatial Gaussian process in more
detail by looking at its distribution at various sites. Now
%
%e13 #&#
\begin{equation}
\label{eqzeta} \vvec{\zeta}|\vec{t},\btheta,\sigma,a_\zeta,r_\zeta
\sim N_n \bigl(\vvec{\mu}_{*}(X),V_{*}(X)
\bigr),
\end{equation}
where
\begin{eqnarray*}
\vvec{\mu}_{*}(X) & = & K_{\zeta}(X,X)^{T} \bigl
\{K_{\zeta}(X,X)+\Sigma_{*} \bigr\}^{-1} \bigl\{
\vec{t}-\vvec {\mu}(\btheta) \bigr\},
\\
V_{*}(X) &=& K_{\zeta}(X,X) - K_{\zeta}(X,X)^{T}
\bigl\{K_{\zeta}(X,X)+\Sigma_{*} \bigr\}^{-1}K_{\zeta}(X,X),
\end{eqnarray*}
and so posterior realizations from the spatial process can be obtained
using the MCMC output
$(\btheta^{(j)},\sigma^{(j)},a_\zeta^{(j)},r_\zeta^{(j)})$,
$j=1,\ldots,J$, by simulating from
$\vvec{\zeta}|\bt,\btheta^{(j)},\sigma^{(j)},a_\zeta^{(j)},r_\zeta^{(j)}$,
$j=1,\ldots,J$. For example,\vspace*{1pt} the posterior spatial process at the UK
site Monamore ($5.1^\circ$W, $55.5^\circ$N) has mean 21.7 years
and standard deviation 33.0 years, and this small mean reflects both
the good fit of the wavefront model at this site and at its
neighboring sites; see Figure~\ref{figrcarb}. In contrast, the
French site Greifensee ($8.7^\circ$W, $47.7^\circ$N) has mean -726
years and standard deviation 453~years, and this large mean
highlights that the data from this site is inconsistent with its
neighbors.

%s6.1 #&#
\subsection{Model fit}
\begin{figure}

\includegraphics{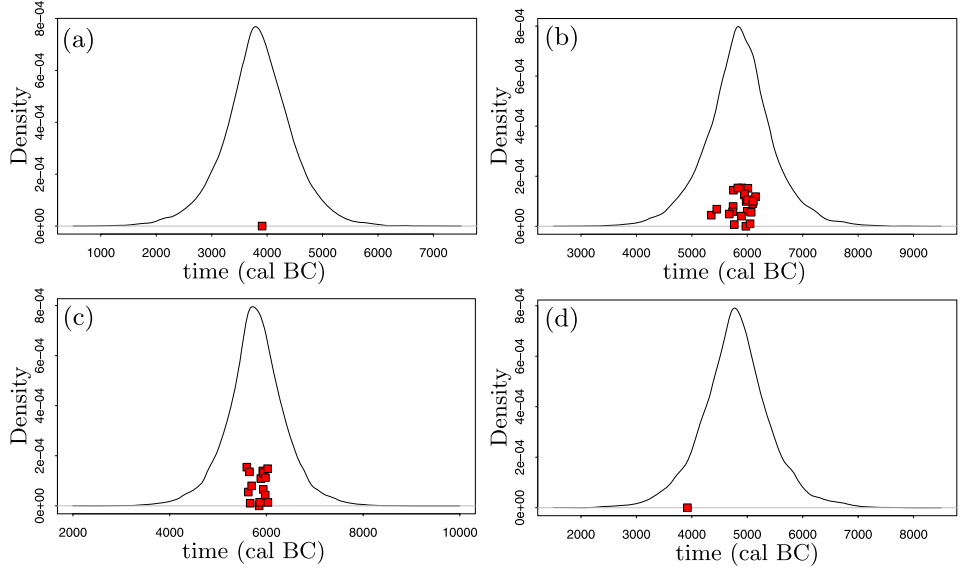}

\caption{$^{14}$C dates (red squares, displaced vertically for
presentation purposes) and their predictive densities (solid)
for four randomly chosen sites: \textup{(a)} Monamore
($m_i=1,\sigma_i=120$), \textup{(b)} Vrsnik-Tarinci
($m_i=22,\sigma_i=13$), \textup{(c)} Galabnik ($m_i=10,\sigma_i=16$) and
\textup{(d)} Greifensee ($m_i=1,\sigma_i=130$).}
\label{pred}
\end{figure}

We use predictive simulations to assess the validity of the
statistical model and thereby the underlying model of the propagating
wavefront. The (joint) posterior predictive distribution of the
arrival times $\vec{t}_{\mathrm{pred}}$ at the $n=302$ radiocarbon sites
can be determined in a similar way to that for the (posterior) spatial
process. First we have
\[
\vec{t}_{\mathrm{pred}}|\btheta,\sigma,\vvec{\zeta}\sim N_n \bigl(
\vvec{\mu}(\btheta)+\vvec{\zeta},\Sigma_{*}(\btheta ) \bigr),
\]
where $\Sigma_{*}(\btheta)=\operatorname{diag}\{V_{i}(\btheta)+\sigma^2_i+\sigma^2\}$
and $\vvec{\zeta}=(\zeta(\bx_{1}),\ldots,\zeta(\bx_{n}))^{T}$
is the spatial Gaussian process evaluated at the sites. We can
integrate over $\vvec{\zeta}$ in closed form using
(\ref{eqzeta}) to obtain
\[
\vec{t}_{\mathrm{pred}}|\bt,\btheta,\sigma,a_\zeta,r_\zeta\sim
N_n \bigl(\vvec{\mu}(\btheta)+\vvec{\mu}_{*}(X),
\Sigma_{*}(\btheta )+V_{*}(X) \bigr)
\]
and thereby determine the predictive arrival time distribution by
simulating realizations from $\vec{t}_{\mathrm{pred}}^{(j)}|\bt,\btheta^{(j)},\sigma^{(j)},a_\zeta^{(j)},r_\zeta^{(j)}$,
$j=1,\ldots,J$. Figure~\ref{pred} shows the radiocarbon dates for
four randomly chosen sites together with their posterior predictive
%
%f3 #&#
distributions determined using this simulation technique. These sites
are at Monamore ($5.1^\circ$W, $55.5^\circ$N), Vrsnik-Tarinci
($22.0^\circ$E, $41.7^\circ$N), Galabnik ($23.1^\circ$E,
$42.4^\circ$N) and Greifensee ($8.7^\circ$W, $47.7^\circ$N).
The $^{14}$C dates cluster very close to the mode of the posterior
densities in three cases and the discrepant site (d) is the one
discussed in the previous section whose data is not consistent with
its neighboring sites. Very encouragingly, we found similar
clustering of site dates around their predictive mode for around 90\%
of the 302 sites. We conclude that the wavefront model (or more
properly, the emulators) provides an adequate description of the data.
%
%f4 #&#
\begin{figure}

\includegraphics{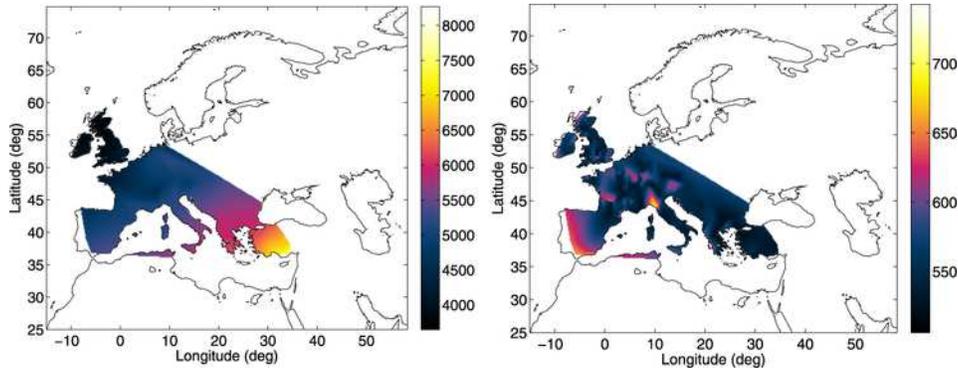}

\caption{The mean (left) and standard deviation (right) of
the posterior predictive distribution of the arrival time.}
\label{smoothedpred}
\end{figure}
Figure~\ref{smoothedpred} illustrates how the wavefront moves
across Europe via a map of the means of the posterior predictive
distributions (Figure~\ref{pred}) generated by interpolating the
mean arrival time at each site onto a regular mesh, using cubic
interpolation with the \textsc{Matlab} routine \texttt{griddata}. In
the same figure we also display the standard deviation of the
predictive distributions interpolated in the same manner,
highlighting areas where there is large uncertainty in the timing of
the wavefront arrival from our model.

%s7 #&#
\section{Discussion}
\label{secdiscuss}
Our main goal in this paper has been to develop generic statistical
tools rather than to propose a more complex model for the spread of
the Neolithic than, for example, \citet{Ackland2007},
\citet{Davison2009} and \citet{Fort2010}. Our approach to inferring
parameters of the Neolithic expansion is based on a direct application
of the wavefront model. To alleviate the computational demands of the
MCMC scheme, we constructed Gaussian process emulators for the arrival
time of the wavefront at each of the 302 radiocarbon-dated sites in
our data set, with the input parameters obtained from the wavefront
model. This approach has been shown to perform well and is likely to
do so for other spatio-temporal models that have complex space--time
dependencies and are slow to simulate.

This paper represents the first attempt of statistical inference for
quantitative models of the Neolithic dispersal based on modern
statistical methods, providing an opportunity for a more detailed and
reliable analysis of the data than ever before. We have improved upon
earlier model fitting for the Neolithic expansion by avoiding the
introduction of a maximum precision of the Early Neolithic $^{14}$C
age determinations at 100--200 years, obtained as the minimum
empirical standard deviation of $^{14}$C dates in large homogeneous
data sets for key Neolithic sites [\citet{DS04,DSGSTZ05,Davison2009}].
In fact, results presented here remain unaffected when such a maximum
precision is introduced. This was not the case with the earlier
analyses, and the robustness of the inference procedure suggested here
is encouraging.

We believe that we have demonstrated the usefulness of our method as a
generic tool and that adopting a statistically rigorous approach
constitutes a significant advance on previous ``wave of advance''
models of the Neolithic. The posterior distribution of the parameters
of our wavefront model ($\nu,V_{\mathrm C},V_{\mathrm R}$) provides clearly
bounded constraints on the human processes underlying the spread of
farming. Also, the posterior distribution of the parameters of our
statistical model ($\sigma,a_{\zeta},r_{\zeta}$) allows a quantitative
assessment of the utility of our wavefront model (including the limits
of applicability of such a large-scale model to smaller regions). The
latter point has important implications for the archaeological
interpretation of the Neolithic transition in Europe and should
contribute to the ongoing debate about the characterization of the
spread as a continuous phenomenon. For example,
\citet{Rowley-Conwy2011} argues that the expansion of farming was
characterized by sporadic events, punctuated by long periods without
outward expansion; Rowley-Conwy includes the spread of the LBK and
Impressed Ware cultures as rapid events of this type. In light of the
difficulties encountered here in modeling these spreads, noted in the
previous section, it would be of interest to consider alternative
models, which are perhaps better able to model the expansion within the relevant
regions.

In terms of future work, the wavefront model can be improved in
several respects. Most importantly, a more detailed description of
regional variations appears to be necessary. In particular, $^{14}$C
data suggest that the accelerated spread along the coastlines is most
pronounced in the western Mediterranean and around the Iberian
Peninsula, but not on the North Sea and Black Sea coasts. The approach
presented here can be readily generalized to include such a variation
and to provide a useful estimate or an upper limit on the coastal
propagation speed in Northern Europe. Also, our understanding of the
Neolithic spread can be enriched by adding other parameters to our
model, such as the starting time and position of the source of the
Neolithic expansion, and even additional sources as suggested by
\citet{Davison2009}, based on the $^{14}$C evidence and corroborated
by \citet{MMHJ09} using archaeobotanical evidence. Another major
improvement would be to use the published $^{14}$C dates directly for
inference, rather than combining them at each site as in
(\ref{eqtisigmai}). Model fits to the original radiocarbon data
without preliminary selection, unavoidably involving subjective and
possibly poorly justified criteria, have never been attempted in the
studies of the Neolithic [\citet{HS04,St10}].

A perennial question in the ongoing debate about the nature of the
Neolithic is the mechanism of the spread of the agropastoral
lifestyle. The wavefront model used here applies equally well to
demic spread, involving migration of people, and to cultural
diffusion, that is, the transition to farming due to indigenous
adaptation and knowledge transfer [\citet{zvelebil1998}]. However,
cultural diffusion may be modeled more explicitly, in terms of
multiple, mutually-interacting populations; see, for example,
\citet{Aoki1996}, \citet{Ackland2007} and \citet{Patterson2010}. We
intend to explore the fit of some of these other models and to
determine how plausible these different models are as explanations of
the radiocarbon data.

\begin{appendix}\label{app}
\section*{Appendix}
%s7.1 #&#
\subsection{Experimental design}
\label{secexptdesign}
In order to fit the emulators, we first need to obtain the training
data by running the wavefront model at a suitable choice of
$\btheta$-values. Computing resources available to us allowed a
$p=200$-point design. Although using a regular lattice design is
appealingly simple, it is not particularly efficient. Instead, we
adopt a more commonly used design for fitting Gaussian processes,
namely, the Latin Hypercube Design (LHD) [\citet{McKay1979}]. Designs
of this class are space-filling in that they distribute points within
a hypercube in parameter space more efficiently than a lattice design.
Our 200-point LHD was constructed using the \texttt{maximin} option in
the \textsc{Matlab} routine \texttt{lhsdesign}. We set the lower
bounds of the hypercube to be the origin and used the upper one
percentiles of the prior distribution as its upper bounds. The routine
then simulated 5000 LHDs and took the design which maximized the
minimum distance between design points. We then used a sequential
strategy in which we repeatedly ran the inference algorithm (described
in the following section), used the results to determine a
conservative estimate of a hypercube containing all points in the MCMC
output (and therefore plausibly containing all of the posterior
density), and then generated another LHD. The final LHD used for
inferences on $(\nu,V_\C,V_\R)^T$ in this paper is contained within
the hypercube $(0,120)\times(0,3)\times(0,2)$.

%s7.2 #&#
\subsection{Fitting the emulator}
\label{secemFit}
\begin{figure}

\includegraphics{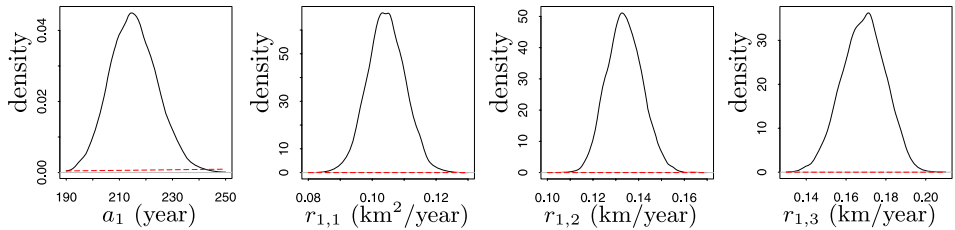}

\caption{Solid (black) curves: the marginal posterior densities of
the emulator's hyperparameters $a_1$ and $\vec{r}_1$ (from
left to right) for the Achilleion site. Dashed (red) curves:
the corresponding prior probability densities.}
\label{fighyperparam}
\end{figure}

At site $i$, this essentially requires the determination of the
posterior distribution of the emulator's hyperparameters using the
training data $\vvec{\tau}_{i}(\Theta)$ from the wavefront model
evaluated at the emulator's design points. We adopt fairly diffuse
independent log-normal priors for the GP hyperparameters, with
$a_i\sim\operatorname{LN}(6.3,0.5)$ and $r_{i,j}\sim\operatorname{LN}(9,3)$. Then
their (joint) posterior density is
\[
\pi(a_{i},\vec{r}_{i}|\vvec{
\tau}_{i})\propto\pi(a_{i})\pi (r_{i,1})
\pi(r_{i,2})\pi(r_{i,3})\pi(\vvec{\tau}_{i}|a_{i},
\vec{r}_{i}),
\]
where
\[
\pi(\vvec{\tau}_{i}|a_{i},\vec{r}_{i})\propto
\bigl|K_{i}(a_{i},\vec{r}_{i})\bigr|^{-1/2}\exp
\bigl\{-\tfrac{1}{2}(\vvec {\tau}_{i}-\vec{m}_{i})^T
K_{i}(a_{i},\vec{r}_{i})^{-1}(\vvec{
\tau }_{i}-\vec{m}_{i}) \bigr\}.
\]
For notational simplicity, we have dropped the explicit dependence of
the training data $\vvec{\tau}_{i}$, the mean $\vec{m}_{i}$ and the
covariance matrix $K_{i}$ on the training points $\Theta$. It is
fairly straightforward to implement an MCMC scheme to simulate
realizations from this posterior distribution. We used a
Metropolis--Hastings scheme in which proposals for each hyperparameter
were made on a log-scale using (univariate) normal random walks. A
suitable magnitude of the innovations (their variance) was adopted
after a series of short trial runs. For each site-specific emulator,
the MCMC scheme was run for 500K iterations after a burn-in of 50K
iterations and the output thinned by taking every 50th iterate,
giving a total of 10K (almost uncorrelated) iterates to be used for
inference. Figure~\ref{fighyperparam} shows the marginal posterior
densities of the emulator hyperparameters for site 1, the Achilleion
%
%f5 #&#
site, located at latitude $39.2^\circ$N, longitude $22.38^\circ$E.
Note that these posterior densities have relatively small variability
and are robust to the parameter choice for the prior of the
hyperparameters, and both these features are typical of all the sites
in our data set.

%s7.3 #&#
\subsection{Emulator validation}
\label{secemVal}
The accuracy of our fitted emulators as an approximation to the
wavefront model can be assessed by using a variety of graphical and
numerical methods. Our assessment was based on the fit at a set of
parameter values that were not used to fit the emulator: we used a
$p^{*}=100$-point Latin hypercube design
$\Theta^{*}=(\btheta^{*}_{1},\ldots,\btheta^{*}_{p^{*}})^{T}$.
First, for each site $i$, we obtained the arrival times
$\vvec{\tau}_{i}=(\tau_i(\btheta_{1}^{*}),\ldots,\tau_i(\btheta_{p^*}^{*}))^T$
from the propagating front model. We then obtained realizations of the
emulator at these design points which properly accounted for the
uncertainty in the hyperparameters by using (\ref{emPred}) and the
realizations from the MCMC fit of the emulator. Plots of the
emulator's mean and 95\% credible region showed that the emulators
provided a reasonably accurate fit to the wavefront output. We note
that these plots were very similar to those determined by fixing the
hyperparameters at their posterior means.

\begin{figure}

\includegraphics{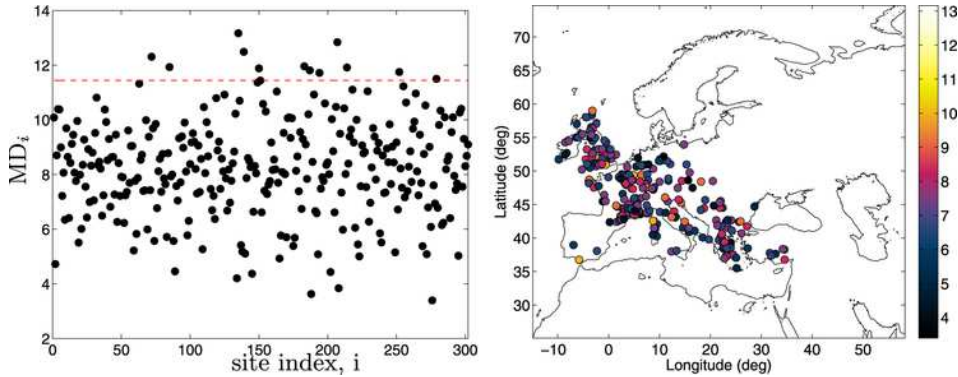}

\caption{Diagnostic plots assessing the fit of the emulator to the
wavefront model. Left panel: the Mahalanobis distance
$\mathrm{MD}_i$ for each site, together with its distribution's upper
$95\%$ point (dashed red line at $\mathrm{MD}_i=11.44$). Right
panel: the spatial distribution of sites colored according to
the Mahalanobis distance.}
\label{figemulatortest}
\end{figure}

We also assessed the fit of our emulators by using a numerical
statistic taken from \citet{Bastos2009} which compares the emulator
output with that from the wavefront model allowing for the
site-specific accuracy and correlation between design points. This
statistic is the site-specific Mahalanobis distance MD$_i$,
$i=1,\ldots,n$, where
\[
\operatorname{MD}_i^2=\bigl(\vvec{\tau}_{i}-
\vvec{\mu_i}\bigl(\Theta^*\bigr)\bigr)^{T}V_{i}
\bigl(\Theta^{*}\bigr)^{-1} \bigl(\vvec{\tau}_{i}-
\vvec{\mu_i}\bigl(\Theta^*\bigr)\bigr),
\]
which has a scaled $F$-distribution in the case of the GP emulator
with fixed hyperparameters, with $\operatorname{MD}_i^2\sim
p^*(p-5)F_{p^{*}, p-3}/(p-3)$. In our calculations we have fixed the
hyperparameters at their posterior means.
Figure~\ref{figemulatortest} shows the Mahalanobis distance at each
%
%f6 #&#
site, together with the upper $95\%$ point of its distribution, and
the spatial distribution of the Mahalanobis distance. The figure
confirms that the emulators provide a reasonable fit throughout the
design space and that there is no obvious spatial dependence in the
fit of the emulators. We note that these diagnostics gave similar
conclusions for all values in the posterior sample of
$(a_i,\vec{r}_i)$ and for different LHDs without any systematic
site-specific biases.

%f7 #&#
\begin{figure}

\includegraphics{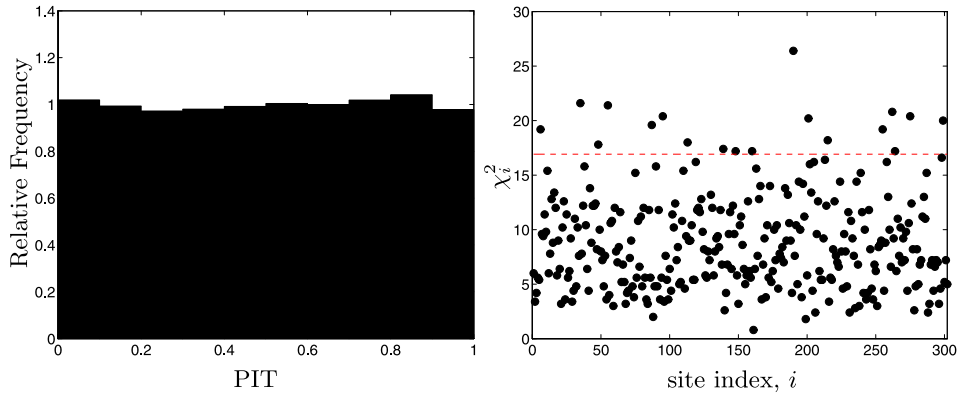}

\caption{Analysis of probability integral transform (PIT) values,
calculated for each site from 100 emulated arrival times
$\vvec{\tau}_{i}^{*}$. Left panel: a histogram of all PIT values
from each of the 302 sites. Right panel: the $\chi^2$ statistic,
computed at each site, assessing the goodness of fit of the PIT
values to a uniform distribution. The 95\% point of the
$\chi_{9}^2$ distribution is plotted as a dashed line.}
\label{figPITtest}
\end{figure}

Finally, we have assessed the distributional fit of the emulators by
calculating the probability integral transform (PIT) values
[\citet{Gneitingetal2007}] for each site at each of the 100 points in
the design. Again, we fix the hyperparameters at their posterior
means. A well fitting emulator should produce values from the
standard uniform distribution. A~histogram of the PIT values over all
sites is shown in Figure~\ref{figPITtest}. The figure also shows a
summary of how closely the PIT values from each site resemble the
standard uniform\vadjust{\goodbreak} distribution. Here we use a Pearson $\chi^2$
statistic after binning the data into ten equal sized bins and so also
show the 95\% point of the $\chi_{9}^2$ distribution. This figure
confirms that the emulator fits are satisfactory.

%s7.4 #&#
\subsection{Allowing for hyperparameter uncertainty}
\label{sechyperuncer}
In Section~\ref{secinference} we describe how to obtain realizations
from the posterior distribution (\ref{jpStatM}) when fixing the
emulator hyperparameters at their posterior mean. Here we extend the
method to take account of hyperparameter uncertainty by using their
realizations
$\{(a_i^{(j)},\vec{r}_i^{(j)});j=1,\ldots,n_e,i=1,\ldots,n\}$ obtained
when fitting the emulators.

We first describe an MCMC scheme in which the emulators themselves are
marginalized over hyperparameter uncertainty. In this case, the
marginal distribution of $\tau_i^*(\btheta)$ can be well approximated
by an equally weighted normal mixture, namely,
\[
\tau_i^*(\btheta)\sim\sum_{j=1}^{n_e}
\frac{1}{n_e} N\bigl(\mu_i\bigl(\btheta;a_i^{(j)},
\vec{r}_i^{(j)}\bigr),V_i\bigl(
\btheta;a_i^{(j)},\vec {r}_i^{(j)}
\bigr)\bigr),
\]
where $\mu_i(\cdot;a,\vec{r})$ and $V_i(\cdot;a,\vec{r})$ are
given by
(\ref{emMean}) and (\ref{emVar}). Therefore, the (emulator)
likelihood in (\ref{jpStatM}) can be written as
\[
\pi_{e}(\bt|\btheta,\sigma,a_\zeta,r_\zeta)=\sum
_{j=1}^{n_e}\frac{1}{n_e}
\phi_n\bigl(\bt|\vvec{\mu}\bigl(\btheta;\vec{a}^{(j)},
\vec{r}^{(j)}\bigr),\Sigma \bigl(\btheta;\vec{a}^{(j)},
\vec{r}^{(j)}\bigr)\bigr),
\]
where $\phi_n(\cdot|\vvec{\mu},\Sigma)$ is the density of an
$n$-dimensional normal distribution with mean $\vvec{\mu}$ and
covariance matrix $\Sigma$, $\vvec{\mu}(\btheta;\vec{a},\vec{r})$ has
$i$th element $\mu_{i}(\btheta;a_i,\vec{r}_i)$ and, to allow for the
emulator uncertainty, we redefine
$\Sigma(\btheta;\vec{a},\vec{r})=K_{\zeta}(X,X)+\operatorname{diag}\{
V_{i}(\btheta;a_i,\vec{r}_i)+\sigma^2_i+\sigma^2\}$,
where $X=(\bx_1,\ldots,\bx_n)^T$. Sampling from the posterior
distribution when using this (emulator) likelihood can be achieved by
using the same methods as described in Section~\ref{secinference}.
However, a major drawback of using this scheme is the time taken to
calculate the (emulator) likelihood at each iteration. Therefore, we
prefer to use an MCMC scheme which uses a joint update for the model
parameters and the hyperparameters and has (\ref{jpStatM}) as a
marginal distribution. Specifically, in the joint update, the value
proposed for the hyperparameters is sampled from the posterior
realizations described above. Although this scheme has a much larger
state space, each MCMC iteration is considerably faster than in the
marginal scheme and we have found it to have good convergence
properties.\looseness=1

%f8 #&#
\begin{figure}

\includegraphics{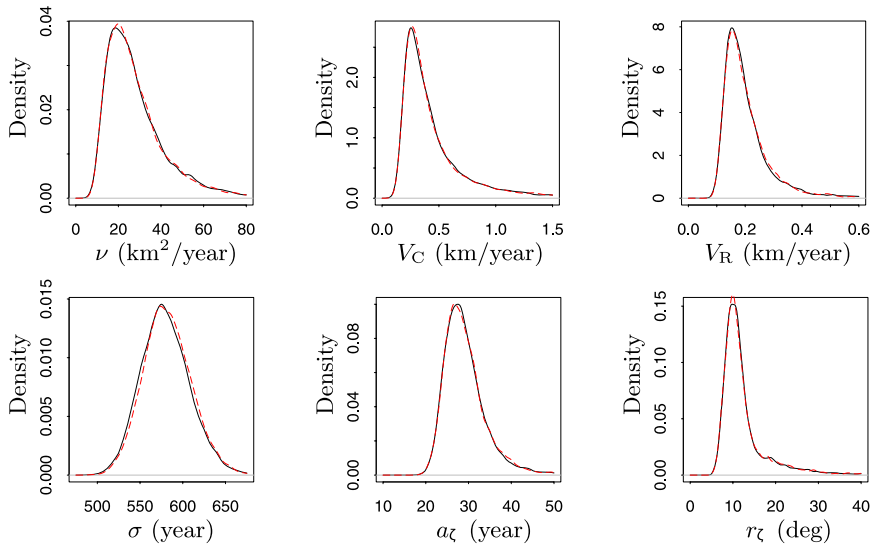}

\caption{Marginal posterior densities for the wavefront model
parameters $(\nu,V_\C,V_\R)$ and for the statistical model
parameters $(\sigma,a_\zeta,r_\zeta)$ allowing for uncertainty
in the hyperparameters (black curves) and for fixed
hyperparameters (dashed red curves).}
\label{figpostplotsnew}
\end{figure}

We ran the MCMC scheme for 1M iterations (after a burn-in of 100K
iterations), and then thinned the output taking every 100th iterate to
leave an effectively uncorrelated sample of 10K draws from the
posterior densities. Figure~\ref{figpostplotsnew} shows the
marginal posterior distributions allowing for emulator hyperparameter
uncertainty and for fixed hyperparameters. We note that it is
difficult to distinguish between these cases. Similar conclusions were
found (when comparing fixed with uncertain hyperparameters) for the
site-specific posterior distribution of the spatial
process $\vvec{\zeta}|\bt$ and for the predictive distributions
$t_{\mathrm{pred}}|\bt$.
\end{appendix}

\section*{Acknowledgments}

The authors would like to thank an Editor and three anonymous referees
for their constructive comments on an earlier draft.

\begin{supplement}%[id=suppA]
\stitle{Data}
\slink[doi]{10.1214/12-AOAS579SUPP} %[doi,text={...}] - jei reikia
%suskaldyti doi
\slink[url]{http://lib.stat.cmu.edu/aoas/579/supplement.zip}
\sdatatype{.zip}
\sdescription{The zip file contains the raw data (xls file) and a
description of the raw data (pdf file).}
\end{supplement}

% imsref loaded by lrinkeviciute, 2012-08-20 11:17:31
% imsref loaded by lrinkeviciute, 2012-08-21 13:01:58
% imsref loaded by lrinkeviciute, 2012-08-21 13:03:00

\printaddresses


\begin{thebibliography}{44}
% BibTex style file: ims.bst, 2012-08-16
% Default style options (sort=0,type=number).
% Used options (sort=1,type=nameyear).

%b1 ###
\bibitem[\protect\citeauthoryear{Ackland et~al.}{2007}]{Ackland2007}
\begin{barticle}[pbm]
\bauthor{\bsnm{Ackland},~\bfnm{Graeme~J.}\binits{G.~J.}},
  \bauthor{\bsnm{Signitzer},~\bfnm{Markus}\binits{M.}},
  \bauthor{\bsnm{Stratford},~\bfnm{Kevin}\binits{K.}} \AND
  \bauthor{\bsnm{Cohen},~\bfnm{Morrel~H.}\binits{M.~H.}}
(\byear{2007}).
\btitle{Cultural hitchhiking on the wave of advance of beneficial
  technologies}.
\bjournal{Proc. Natl. Acad. Sci. USA}
\bvolume{104}
\bpages{8714--8719}.
\bid{doi={10.1073/pnas.0702469104}, issn={0027-8424}, pii={0702469104},
  pmcid={1885568}, pmid={17517663}}
\bptok{imsref}%
\end{barticle}
\endbibitem

%b2 ###
\bibitem[\protect\citeauthoryear{Aitken}{1990}]{A90}
\begin{bbook}[author]
\bauthor{\bsnm{Aitken},~\bfnm{M.}\binits{M.}}
(\byear{1990}).
\btitle{Science-Based Dating in Archaeology}.
\bpublisher{Longman}, \blocation{Harlow}.
\bptok{imsref}%
\end{bbook}
\endbibitem

%b3 ###
\bibitem[\protect\citeauthoryear{Ammerman and
  Cavalli-Sforza}{1971}]{Ammerman1971}
\begin{barticle}[author]
\bauthor{\bsnm{Ammerman},~\bfnm{A.~J.}\binits{A.~J.}} \AND
  \bauthor{\bsnm{Cavalli-Sforza},~\bfnm{L.~L.}\binits{L.~L.}}
(\byear{1971}).
\btitle{Measuring the rate of spread of early farming in Europe}.
\bjournal{Man}
\bvolume{6}
\bpages{674--688}.
\bptok{imsref}%
\end{barticle}
\endbibitem

%b4 ###
\bibitem[\protect\citeauthoryear{Ammerman and
  Cavalli-Sforza}{1973}]{Ammerman1973}
\begin{bincollection}[author]
\bauthor{\bsnm{Ammerman},~\bfnm{A.~J.}\binits{A.~J.}} \AND
  \bauthor{\bsnm{Cavalli-Sforza},~\bfnm{L.~L.}\binits{L.~L.}}
(\byear{1973}).
\btitle{A population model for the diffusion of early farming in Europe}.
In \bbooktitle{The Explanation of Culture Change}
(\beditor{\bfnm{Colin}\binits{C.}~\bsnm{Renfrew}}, ed.)
\bpages{343--357}.
\bpublisher{Duckworth}, \blocation{London}.
\bptok{imsref}%
\end{bincollection}
\endbibitem

%b5 ###
\bibitem[\protect\citeauthoryear{Aoki, Shida and Shigesada}{1996}]{Aoki1996}
\begin{barticle}[author]
\bauthor{\bsnm{Aoki},~\bfnm{Kenichi}\binits{K.}},
  \bauthor{\bsnm{Shida},~\bfnm{Mitsuo}\binits{M.}} \AND
  \bauthor{\bsnm{Shigesada},~\bfnm{Nanako}\binits{N.}}
(\byear{1996}).
\btitle{Travelling wave solutions for the spread of farmers into a region
  occupied by hunter-gatherers}.
\bjournal{Theor. Popul. Biol.}
\bvolume{50}
\bpages{1--17}.
\bptok{imsref}%
\end{barticle}
\endbibitem

%b6 ###
\bibitem[\protect\citeauthoryear{Baggaley et~al.}{2009}]{Baggaley2009}
\begin{barticle}[author]
\bauthor{\bsnm{Baggaley},~\bfnm{Andrew~W.}\binits{A.~W.}},
  \bauthor{\bsnm{Barenghi},~\bfnm{Carlo~F.}\binits{C.~F.}},
  \bauthor{\bsnm{Shukurov},~\bfnm{Anvar}\binits{A.}} \AND
  \bauthor{\bsnm{Subramanian},~\bfnm{Kandaswamy}\binits{K.}}
(\byear{2009}).
\btitle{Reconnecting flux-rope dynamo}.
\bjournal{Phys. Rev. E}
\bvolume{80}
\bpages{055301}.
\bptok{imsref}%
\end{barticle}
\endbibitem

%b7 ###
\bibitem[\protect\citeauthoryear{Baggaley et~al.}{2013}]{Baggaley2013}
\begin{bmisc}[author]
\bauthor{\bsnm{Baggaley},~\bfnm{Andrew~W.}\binits{A.~W.}},
  \bauthor{\bsnm{Boys},~\bfnm{Richard~J.}\binits{R.~J.}},
  \bauthor{\bsnm{Golightly},~\bfnm{Andrew}\binits{A.}} \AND
  \bauthor{\bsnm{Shukurov},~\bfnm{Anvar}\binits{A.}}
(\byear{2013}).
\bhowpublished{Supplement to ``Inference for population dynamics in the
Neolithic period.'' DOI:\doiurl{10.1214/12-AOAS579SUPP}.}
\bptok{imsref}%
\end{bmisc}
\endbibitem

%b8 ###
\bibitem[\protect\citeauthoryear{Bastos and O'Hagan}{2009}]{Bastos2009}
\begin{barticle}[mr]
\bauthor{\bsnm{Bastos},~\bfnm{Leonardo~S.}\binits{L.~S.}} \AND
  \bauthor{\bsnm{O'Hagan},~\bfnm{Anthony}\binits{A.}}
(\byear{2009}).
\btitle{Diagnostics for {G}aussian process emulators}.
\bjournal{Technometrics}
\bvolume{51}
\bpages{425--438}.
\bid{doi={10.1198/TECH.2009.08019}, issn={0040-1706}, mr={2756478}}
\bptok{imsref}%
\end{barticle}
\endbibitem

%b9 ###
\bibitem[\protect\citeauthoryear{Bayarri et~al.}{2007}]{bayarri07}
\begin{barticle}[mr]
\bauthor{\bsnm{Bayarri},~\bfnm{Maria~J.}\binits{M.~J.}},
  \bauthor{\bsnm{Berger},~\bfnm{James~O.}\binits{J.~O.}},
  \bauthor{\bsnm{Paulo},~\bfnm{Rui}\binits{R.}},
  \bauthor{\bsnm{Sacks},~\bfnm{Jerry}\binits{J.}},
  \bauthor{\bsnm{Cafeo},~\bfnm{John~A.}\binits{J.~A.}},
  \bauthor{\bsnm{Cavendish},~\bfnm{James}\binits{J.}},
  \bauthor{\bsnm{Lin}, \bfnm{Chin-Hsu}\binits{C.-H.}} \AND
  \bauthor{\bsnm{Tu},~\bfnm{Jian}\binits{J.}}
(\byear{2007}).
\btitle{A framework for validation of computer models}.
\bjournal{Technometrics}
\bvolume{49}
\bpages{138--154}.
\bid{doi={10.1198/004017007000000092}, issn={0040-1706}, mr={2380530}}
\bptok{imsref}%
\end{barticle}
\endbibitem

%b10 ###
\bibitem[\protect\citeauthoryear{Birdsell}{1957}]{Birdsell1957}
\begin{barticle}[author]
\bauthor{\bsnm{Birdsell},~\bfnm{J.~B.}\binits{J.~B.}}
(\byear{1957}).
\btitle{Some population problems involving Pleistocene man}.
\bjournal{Cold Spring Harbor Symposium on Quantitative Biology}
\bvolume{22}
\bpages{47--69}.
\bptok{imsref}%
\end{barticle}
\endbibitem

%b11 ###
\bibitem[\protect\citeauthoryear{Blackwell and Buck}{2008}]{BB08}
\begin{barticle}[mr]
\bauthor{\bsnm{Blackwell},~\bfnm{P.~G.}\binits{P.~G.}} \AND
  \bauthor{\bsnm{Buck},~\bfnm{C.~E.}\binits{C.~E.}}
(\byear{2008}).
\btitle{Estimating radiocarbon calibration curves}.
\bjournal{Bayesian Anal.}
\bvolume{3}
\bpages{225--248}.
\bid{issn={1936-0975}, mr={2407423}}
\bptok{imsref}%
\end{barticle}
\endbibitem

%b12 ###
\bibitem[\protect\citeauthoryear{Bocquet-Appel et~al.}{2009}]{BNLK09}
\begin{barticle}[author]
\bauthor{\bsnm{Bocquet-Appel},~\bfnm{Jean-Pierr{\'e}}\binits{J.-P.}},
  \bauthor{\bsnm{Naji},~\bfnm{Stephan}\binits{S.}},
  \bauthor{\bsnm{Linden},~\bfnm{Marc~Vander}\binits{M.~V.}} \AND
  \bauthor{\bsnm{Kozlowski},~\bfnm{Janusz~K.}\binits{J.~K.}}
(\byear{2009}).
\btitle{Detection of diffusion and contact zones of early farming in Europe
  from the space--time distribution of 14C dates}.
\bjournal{J. Archeo. Sci.}
\bvolume{36}
\bpages{807--820}.
\bptok{imsref}%
\end{barticle}
\endbibitem

%b13 ###
\bibitem[\protect\citeauthoryear{Buck et~al.}{1991}]{BKLS91}
\begin{barticle}[author]
\bauthor{\bsnm{Buck},~\bfnm{C.~E.}\binits{C.~E.}},
  \bauthor{\bsnm{Kenworthy},~\bfnm{J.~B.}\binits{J.~B.}},
  \bauthor{\bsnm{Litton},~\bfnm{C.~D.}\binits{C.~D.}} \AND
  \bauthor{\bsnm{Smith},~\bfnm{A.~F.~M.}\binits{A.~F.~M.}}
(\byear{1991}).
\btitle{Combining archaeological and radiocarbon information: A Bayesian
  approach to calibration}.
\bjournal{Antiquity}
\bvolume{65}
\bpages{808--821}.
\bptok{imsref}%
\end{barticle}
\endbibitem

%b14 ###
\bibitem[\protect\citeauthoryear{Burger and Thomas}{2011}]{BurgerThomas2011}
\begin{bincollection}[author]
\bauthor{\bsnm{Burger},~\bfnm{Joachim}\binits{J.}} \AND
  \bauthor{\bsnm{Thomas},~\bfnm{Mark~G.}\binits{M.~G.}}
(\byear{2011}).
\btitle{The palaeopopulationgenetics of humans, cattle and diarying in
  neolithic Europe}.
In \bbooktitle{Human Bioarchaeology of the Transition to Agriculture}
(\beditor{\bfnm{Ron}\binits{R.}~\bsnm{Pinhasi}} \AND
  \beditor{\bfnm{Jay~T.}\binits{J.~T.}~\bsnm{Stock}}, eds.)
\bpages{1029--1058}.
\bpublisher{Wiley}, \blocation{Chichester}.
\bptok{imsref}%
\end{bincollection}
\endbibitem

%b15 ###
\bibitem[\protect\citeauthoryear{Davison et~al.}{2006}]{Davison2006}
\begin{barticle}[author]
\bauthor{\bsnm{Davison},~\bfnm{K.}\binits{K.}},
  \bauthor{\bsnm{Dolukhanov},~\bfnm{P.}\binits{P.}},
  \bauthor{\bsnm{Sarson},~\bfnm{G.~R.}\binits{G.~R.}} \AND
  \bauthor{\bsnm{Shukurov},~\bfnm{A.}\binits{A.}}
(\byear{2006}).
\btitle{The role of waterways in the spread of the Neolithic}.
\bjournal{J. Archeo. Sci.}
\bvolume{33}
\bpages{641--652}.
\bptok{imsref}%
\end{barticle}
\endbibitem

%b16 ###
\bibitem[\protect\citeauthoryear{Davison et~al.}{2009}]{Davison2009}
\begin{barticle}[author]
\bauthor{\bsnm{Davison},~\bfnm{K.}\binits{K.}},
  \bauthor{\bsnm{Dolukhanov},~\bfnm{P.~M.}\binits{P.~M.}},
  \bauthor{\bsnm{Sarson},~\bfnm{G.~R.}\binits{G.~R.}},
  \bauthor{\bsnm{Shukurov},~\bfnm{A.}\binits{A.}} \AND
  \bauthor{\bsnm{Zaitseva},~\bfnm{G.~I.}\binits{G.~I.}}
(\byear{2009}).
\btitle{Multiple sources of the European Neolithic: Mathematical modelling
  constrained by radiocarbon dates}.
\bjournal{Quaternary International}
\bvolume{203}
\bpages{10--18}.
\bptok{imsref}%
\end{barticle}
\endbibitem

%b17 ###
\bibitem[\protect\citeauthoryear{Dolukhanov and Shukurov}{2004}]{DS04}
\begin{barticle}[author]
\bauthor{\bsnm{Dolukhanov},~\bfnm{Pavel}\binits{P.}} \AND
  \bauthor{\bsnm{Shukurov},~\bfnm{Anvar}\binits{A.}}
(\byear{2004}).
\btitle{Modelling the Neolithic dispersal in Northern Eurasia}.
\bjournal{Documenta Praehistorica}
\bvolume{31}
\bpages{35--47}.
\bptok{imsref}%
\end{barticle}
\endbibitem

%b18 ###
\bibitem[\protect\citeauthoryear{Dolukhanov et~al.}{2005}]{DSGSTZ05}
\begin{barticle}[author]
\bauthor{\bsnm{Dolukhanov},~\bfnm{Pavel}\binits{P.}},
  \bauthor{\bsnm{Shukurov},~\bfnm{Anvar}\binits{A.}},
  \bauthor{\bsnm{Gronenborn},~\bfnm{Detlef}\binits{D.}},
  \bauthor{\bsnm{Sokoloff},~\bfnm{Dmitry}\binits{D.}},
  \bauthor{\bsnm{Timofeev},~\bfnm{Vladimir}\binits{V.}} \AND
  \bauthor{\bsnm{Zaitseva},~\bfnm{Ganna}\binits{G.}}
(\byear{2005}).
\btitle{The chronology of Neolithic dispersal in Central and Eastern Europe}.
\bjournal{J. Archeo. Sci.}
\bvolume{32}
\bpages{1441--1458}.
\bptok{imsref}%
\end{barticle}
\endbibitem

%b19 ###
\bibitem[\protect\citeauthoryear{Fisher}{1937}]{F37}
\begin{barticle}[author]
\bauthor{\bsnm{Fisher},~\bfnm{R.~A.}\binits{R.~A.}}
(\byear{1937}).
\btitle{The wave of advance of advantageous genes}.
\bjournal{Ann. Eugenics}
\bvolume{7}
\bpages{353--369}.
\bptok{imsref}%
\end{barticle}
\endbibitem

%b20 ###
\bibitem[\protect\citeauthoryear{Fort and Pujol}{2008}]{FP08}
\begin{barticle}[author]
\bauthor{\bsnm{Fort},~\bfnm{Joaquim}\binits{J.}} \AND
  \bauthor{\bsnm{Pujol},~\bfnm{Pujol~Toni}\binits{P.~T.}}
(\byear{2008}).
\btitle{Progress in front propagation research}.
\bjournal{Rep. Progr. Phys.}
\bvolume{71}
\bpages{086001 (41~pp.)}.
\bptok{imsref}%
\end{barticle}
\endbibitem

%b21 ###
\bibitem[\protect\citeauthoryear{Geophysical Data System}{2011}]{geodas}
\begin{bmisc}[author]
\borganization{Geophysical Data System}
(\byear{2011}).
\bhowpublished{Available at
\texttt{\href{http://www.ngdc.noaa.gov/mgg/geodas/geodas.html}{http://www.ngdc.noaa.gov/mgg/geodas/}
\href{http://www.ngdc.noaa.gov/mgg/geodas/geodas.html}{geodas.html}}.}
\bptok{imsref}%
\end{bmisc}
\endbibitem

%b22 ###
\bibitem[\protect\citeauthoryear{Gkiasta et~al.}{2003}]{Gkiasta2003}
\begin{barticle}[author]
\bauthor{\bsnm{Gkiasta},~\bfnm{Marina}\binits{M.}},
  \bauthor{\bsnm{Russell},~\bfnm{Thembi}\binits{T.}},
  \bauthor{\bsnm{Shennan},~\bfnm{Stephen}\binits{S.}} \AND
  \bauthor{\bsnm{Steele},~\bfnm{James}\binits{J.}}
(\byear{2003}).
\btitle{Neolithic transition in Europe: The radiocarbon record revisited}.
\bjournal{Antiquity}
\bvolume{77}
\bpages{45--62}.
\bptok{imsref}%
\end{barticle}
\endbibitem

%b23 ###
\bibitem[\protect\citeauthoryear{Gneiting, Balabdaoui and
  Raftery}{2007}]{Gneitingetal2007}
\begin{barticle}[mr]
\bauthor{\bsnm{Gneiting},~\bfnm{Tilmann}\binits{T.}},
  \bauthor{\bsnm{Balabdaoui},~\bfnm{Fadoua}\binits{F.}} \AND
  \bauthor{\bsnm{Raftery},~\bfnm{Adrian~E.}\binits{A.~E.}}
(\byear{2007}).
\btitle{Probabilistic forecasts, calibration and sharpness}.
\bjournal{J. R. Stat. Soc. Ser. B Stat. Methodol.}
\bvolume{69}
\bpages{243--268}.
\bid{doi={10.1111/j.1467-9868.2007.00587.x}, issn={1369-7412}, mr={2325275}}
\bptok{imsref}%
\end{barticle}
\endbibitem

%b24 ###
\bibitem[\protect\citeauthoryear{Hazelwood and Steele}{2004}]{HS04}
\begin{barticle}[author]
\bauthor{\bsnm{Hazelwood},~\bfnm{Lee}\binits{L.}} \AND
  \bauthor{\bsnm{Steele},~\bfnm{James}\binits{J.}}
(\byear{2004}).
\btitle{Spatial dynamics of human dispersals. Constraints on modelling and
  archaeological validation}.
\bjournal{J. Archeo. Sci.}
\bvolume{31}
\bpages{669--679}.
\bptok{imsref}%
\end{barticle}
\endbibitem

%b25 ###
\bibitem[\protect\citeauthoryear{Henderson et~al.}{2009}]{henderson09}
\begin{barticle}[mr]
\bauthor{\bsnm{Henderson},~\bfnm{Daniel~A.}\binits{D.~A.}},
  \bauthor{\bsnm{Boys},~\bfnm{Richard~J.}\binits{R.~J.}},
  \bauthor{\bsnm{Krishnan},~\bfnm{Kim~J.}\binits{K.~J.}},
  \bauthor{\bsnm{Lawless},~\bfnm{Conor}\binits{C.}} \AND
  \bauthor{\bsnm{Wilkinson},~\bfnm{Darren~J.}\binits{D.~J.}}
(\byear{2009}).
\btitle{Bayesian emulation and calibration of a stochastic computer model of
  mitochondrial {DNA} deletions in substantia nigra neurons}.
\bjournal{J. Amer. Statist. Assoc.}
\bvolume{104}
\bpages{76--87}.
\bid{doi={10.1198/jasa.2009.0005}, issn={0162-1459}, mr={2663034}}
\bptok{imsref}%
\end{barticle}
\endbibitem

%b26 ###
\bibitem[\protect\citeauthoryear{Isern and Fort}{2010}]{Fort2010}
\begin{barticle}[author]
\bauthor{\bsnm{Isern},~\bfnm{N.}\binits{N.}} \AND
  \bauthor{\bsnm{Fort},~\bfnm{J.}\binits{J.}}
(\byear{2010}).
\btitle{Anisotropic dispersion, space competition and the slowdown of the
  Neolithic transition}.
\bjournal{New J. Phys.}
\bvolume{12}
\bpages{123002}.
\bptok{imsref}%
\end{barticle}
\endbibitem

%b27 ###
\bibitem[\protect\citeauthoryear{Kennedy and O'Hagan}{2001}]{kennedy01}
\begin{barticle}[mr]
\bauthor{\bsnm{Kennedy},~\bfnm{Marc~C.}\binits{M.~C.}} \AND
  \bauthor{\bsnm{O'Hagan},~\bfnm{Anthony}\binits{A.}}
(\byear{2001}).
\btitle{Bayesian calibration of computer models}.
\bjournal{J.~R.~Stat. Soc. Ser. B Stat. Methodol.}
\bvolume{63}
\bpages{425--464}.
\bid{doi={10.1111/1467-9868.00294}, issn={1369-7412}, mr={1858398}}
\bptok{imsref}%
\end{barticle}
\endbibitem

%b28 ###
\bibitem[\protect\citeauthoryear{Kolmogorov, Petrovskii and
  Piskunov}{1937}]{Kolmogorov1937}
\begin{barticle}[author]
\bauthor{\bsnm{Kolmogorov},~\bfnm{A.}\binits{A.}},
  \bauthor{\bsnm{Petrovskii},~\bfnm{I.}\binits{I.}} \AND
  \bauthor{\bsnm{Piskunov},~\bfnm{N.}\binits{N.}}
(\byear{1937}).
\btitle{A study of the equation of diffusion with increase in the quantity of
  matter, and its application to a biological problem}.
\bjournal{Byul. Moskovskogo Gos. Univ.}
\bvolume{1}
\bpages{1--25}.
\bptok{imsref}%
\end{barticle}
\endbibitem

%b29 ###
\bibitem[\protect\citeauthoryear{McKay, Beckman and Conover}{1979}]{McKay1979}
\begin{barticle}[mr]
\bauthor{\bsnm{McKay},~\bfnm{M.~D.}\binits{M.~D.}},
  \bauthor{\bsnm{Beckman},~\bfnm{R.~J.}\binits{R.~J.}} \AND
  \bauthor{\bsnm{Conover},~\bfnm{W.~J.}\binits{W.~J.}}
(\byear{1979}).
\btitle{A comparison of three methods for selecting values of input variables
  in the analysis of output from a computer code}.
\bjournal{Technometrics}
\bvolume{21}
\bpages{239--245}.
\bid{doi={10.2307/1268522}, issn={0040-1706}, mr={0533252}}
\bptok{imsref}%
\end{barticle}
\endbibitem

%b30 ###
\bibitem[\protect\citeauthoryear{Motuzaite-Matuzeviciute, Hunt and
  Jones}{2009}]{MMHJ09}
\begin{bincollection}[author]
\bauthor{\bsnm{Motuzaite-Matuzeviciute},~\bfnm{G.}\binits{G.}},
  \bauthor{\bsnm{Hunt},~\bfnm{H.~V.}\binits{H.~V.}} \AND
  \bauthor{\bsnm{Jones},~\bfnm{M.~K.}\binits{M.~K.}}
(\byear{2009}).
\btitle{Multiple sources for Neolithic European agriculture: Geographical
  origins of early domesticates in Moldova and Ukraine}.
In \bbooktitle{The East European Plain on the Eve of Agriculture}
(\beditor{\bfnm{Pavel~M.}\binits{P.~M.}~\bsnm{Dolukhanov}},
  \beditor{\bfnm{Graeme~R.}\binits{G.~R.}~\bsnm{Sarson}} \AND
  \beditor{\bfnm{Anvar}\binits{A.}~\bsnm{Shukurov}}, eds.).
\bseries{British Archaeological Reports, International Series}
\bvolume{1964}
\bpages{53--64}.
\bpublisher{Archaeopress}, \blocation{Oxford}.
\bptok{imsref}%
\end{bincollection}
\endbibitem

%b31 ###
\bibitem[\protect\citeauthoryear{Pape}{2004}]{CIA}
\begin{bmisc}[author]
\bauthor{\bsnm{Pape},~\bfnm{D.}\binits{D.}}
(\byear{2004}).
\bhowpublished{World DataBank II.
Available at
\texttt{\href{http://www.evl.uic.edu/pape/data/WDB/}{http://www.evl.uic.edu/pape/data/}
\href{http://www.evl.uic.edu/pape/data/WDB/}{WDB/}}.}
\bptok{imsref}%
\end{bmisc}
\endbibitem

%b32 ###
\bibitem[\protect\citeauthoryear{Patterson et~al.}{2010}]{Patterson2010}
\begin{barticle}[author]
\bauthor{\bsnm{Patterson},~\bfnm{M.~A.}\binits{M.~A.}},
  \bauthor{\bsnm{Sarson},~\bfnm{G.~R.}\binits{G.~R.}},
  \bauthor{\bsnm{Sarson},~\bfnm{H.~C.}\binits{H.~C.}} \AND
  \bauthor{\bsnm{Shukurov},~\bfnm{A.}\binits{A.}}
(\byear{2010}).
\btitle{Modelling the Neolithic transition in a heterogeneous environment}.
\bjournal{J. Archeo. Sci.}
\bvolume{37}
\bpages{2929--2937}.
\bptok{imsref}%
\end{barticle}
\endbibitem

%b33 ###
\bibitem[\protect\citeauthoryear{Pettitt et~al.}{2003}]{Pettitt2003}
\begin{barticle}[author]
\bauthor{\bsnm{Pettitt},~\bfnm{P.~B.}\binits{P.~B.}},
  \bauthor{\bsnm{Davies},~\bfnm{W.}\binits{W.}},
  \bauthor{\bsnm{Gamble},~\bfnm{C.~S.}\binits{C.~S.}} \AND
  \bauthor{\bsnm{Richards},~\bfnm{M.~B.}\binits{M.~B.}}
(\byear{2003}).
\btitle{Palaeolithic radiocarbon chronology: Quantifying our confidence beyond
  two half-lives}.
\bjournal{J. Archaeo. Sci.}
\bvolume{30}
\bpages{1685--1693}.
\bptok{imsref}%
\end{barticle}
\endbibitem

%b34 ###
\bibitem[\protect\citeauthoryear{Rasmussen and Williams}{2006}]{Rasmussen2006}
\begin{bbook}[mr]
\bauthor{\bsnm{Rasmussen},~\bfnm{Carl~Edward}\binits{C.~E.}} \AND
  \bauthor{\bsnm{Williams},~\bfnm{Christopher K.~I.}\binits{C.~K.~I.}}
(\byear{2006}).
\btitle{Gaussian Processes for Machine Learning}.
\bpublisher{MIT Press}, \blocation{Cambridge, MA}.
\bid{mr={2514435}}
\bptok{imsref}%
\end{bbook}
\endbibitem

%b35 ###
\bibitem[\protect\citeauthoryear{Reimer et~al.}{2004}]{intcal04}
\begin{barticle}[author]
\bauthor{\bsnm{Reimer},~\bfnm{P.~J.}\binits{P.~J.}},
  \bauthor{\bsnm{Baillie},~\bfnm{M.}\binits{M.}},
  \bauthor{\bsnm{Bard},~\bfnm{E.}\binits{E.}},
  \bauthor{\bsnm{Bayliss},~\bfnm{A.}\binits{A.}},
  \bauthor{\bsnm{Beck},~\bfnm{J.}\binits{J.}},
  \bauthor{\bsnm{Bertrand},~\bfnm{C.}\binits{C.}},
  \bauthor{\bsnm{Blackwell},~\bfnm{P.}\binits{P.}},
  \bauthor{\bsnm{Buck},~\bfnm{C.}\binits{C.}},
  \bauthor{\bsnm{Burr},~\bfnm{G.}\binits{G.}},
  \bauthor{\bsnm{Cutler},~\bfnm{K.}\binits{K.}},
  \bauthor{\bsnm{Damon},~\bfnm{P.}\binits{P.}},
  \bauthor{\bsnm{Edwards},~\bfnm{R.}\binits{R.}},
  \bauthor{\bsnm{Fairbanks},~\bfnm{R.}\binits{R.}},
  \bauthor{\bsnm{Friedrich},~\bfnm{M.}\binits{M.}},
  \bauthor{\bsnm{Guilderson},~\bfnm{T.}\binits{T.}},
  \bauthor{\bsnm{Hughen},~\bfnm{K.}\binits{K.}},
  \bauthor{\bsnm{Kromer},~\bfnm{B.}\binits{B.}},
  \bauthor{\bsnm{McCormac},~\bfnm{F.}\binits{F.}},
  \bauthor{\bsnm{Manning},~\bfnm{S.}\binits{S.}},
  \bauthor{\bsnm{Bronk~Ramsey},~\bfnm{C.}\binits{C.}},
  \bauthor{\bsnm{Reimer},~\bfnm{R.}\binits{R.}},
  \bauthor{\bsnm{Remmele},~\bfnm{S.}\binits{S.}},
  \bauthor{\bsnm{Southon},~\bfnm{J.}\binits{J.}},
  \bauthor{\bsnm{Stuiver},~\bfnm{M.}\binits{M.}},
  \bauthor{\bsnm{Talamo},~\bfnm{S.}\binits{S.}},
  \bauthor{\bsnm{Taylor},~\bfnm{F.}\binits{F.}}, \bauthor{\bparticle{van~der}
  \bsnm{Plicht},~\bfnm{J.}\binits{J.}} \AND
  \bauthor{\bsnm{Weyhenmeyer},~\bfnm{C.}\binits{C.}}
(\byear{2004}).
\btitle{IntCal04 terrestrial radiocarbon age calibration}.
\bjournal{Radiocarbon}
\bvolume{46}
\bpages{1029--1058}.
\bptok{imsref}%
\end{barticle}
\endbibitem

%b36 ###
\bibitem[\protect\citeauthoryear{Rowley-Conwy}{2011}]{Rowley-Conwy2011}
\begin{barticle}[author]
\bauthor{\bsnm{Rowley-Conwy},~\bfnm{Peter}\binits{P.}}
(\byear{2011}).
\btitle{Westward ho! The spread of agriculture from Central Europe to the
  Atlantic}.
\bjournal{Current Anthropology}
\bvolume{52}
\bpages{S431--S451}.
\bptok{imsref}%
\end{barticle}
\endbibitem

%b37 ###
\bibitem[\protect\citeauthoryear{Santner, Williams and
  Notz}{2003}]{SantnerWN03}
\begin{bbook}[mr]
\bauthor{\bsnm{Santner},~\bfnm{Thomas~J.}\binits{T.~J.}},
  \bauthor{\bsnm{Williams},~\bfnm{Brian~J.}\binits{B.~J.}} \AND
  \bauthor{\bsnm{Notz},~\bfnm{William~I.}\binits{W.~I.}}
(\byear{2003}).
\btitle{The Design and Analysis of Computer Experiments}.
\bpublisher{Springer}, \blocation{New York}.
\bid{mr={2160708}}
\bptok{imsref}%
\end{bbook}
\endbibitem

%b38 ###
\bibitem[\protect\citeauthoryear{Scott, Cook and Naysmith}{2007}]{SCN07}
\begin{barticle}[author]
\bauthor{\bsnm{Scott},~\bfnm{E.~Marian}\binits{E.~M.}},
  \bauthor{\bsnm{Cook},~\bfnm{Gordon~T.}\binits{G.~T.}} \AND
  \bauthor{\bsnm{Naysmith},~\bfnm{Philip}\binits{P.}}
(\byear{2007}).
\btitle{Error and uncertainty in radiocarbon measurements}.
\bjournal{Radiocarbon}
\bvolume{49}
\bpages{427--440}.
\bptok{imsref}%
\end{barticle}
\endbibitem

%b39 ###
\bibitem[\protect\citeauthoryear{Steele}{2010}]{St10}
\begin{barticle}[author]
\bauthor{\bsnm{Steele},~\bfnm{James}\binits{J.}}
(\byear{2010}).
\btitle{Radiocarbon dates as data: Quantitative strategies for estimating
  colonization front speeds and event densities}.
\bjournal{J. Archeo. Sci.}
\bvolume{37}
\bpages{2017--2030}.
\bptok{imsref}%
\end{barticle}
\endbibitem

%b40 ###
\bibitem[\protect\citeauthoryear{Steele, Adams and Sluckin}{1998}]{Steele1998}
\begin{barticle}[author]
\bauthor{\bsnm{Steele},~\bfnm{James}\binits{J.}},
  \bauthor{\bsnm{Adams},~\bfnm{Jonathan}\binits{J.}} \AND
  \bauthor{\bsnm{Sluckin},~\bfnm{Tim}\binits{T.}}
(\byear{1998}).
\btitle{Modelling Paleoindian dispersals}.
\bjournal{World Archaeology}
\bvolume{30}
\bpages{286--305}.
\bptok{imsref}%
\end{barticle}
\endbibitem

%b41 ###
\bibitem[\protect\citeauthoryear{Whittle}{1996}]{Whittle1996}
\begin{bbook}[author]
\bauthor{\bsnm{Whittle},~\bfnm{Alasdair}\binits{A.}}
(\byear{1996}).
\btitle{Europe in the Neolithic: The Creation of New Worlds}.
\bpublisher{Cambridge Univ. Press}, \blocation{Cambridge}.
\bptok{imsref}%
\end{bbook}
\endbibitem

%b42 ###
\bibitem[\protect\citeauthoryear{Zilh{\~a}o}{2001}]{Zilhao2001}
\begin{barticle}[author]
\bauthor{\bsnm{Zilh{\~a}o},~\bfnm{Joao}\binits{J.}}
(\byear{2001}).
\btitle{Radiocarbon evidence for maritime pioneer colonization at the origins
  of farming in west Mediterranean Europe}.
\bjournal{Proc. Natl. Acad. Sci. USA}
\bvolume{98}
\bpages{14180--14185}.
\bptok{imsref}%
\end{barticle}
\endbibitem

%b43 ###
\bibitem[\protect\citeauthoryear{Zvelebil, Doman{\'{\i}}-ska and
  Dennell}{1998}]{zvelebil1998}
\begin{bbook}[author]
\bauthor{\bsnm{Zvelebil},~\bfnm{M.}\binits{M.}},
  \bauthor{\bsnm{Doman{\'{\i}}-ska},~\bfnm{L.}\binits{L.}} \AND
  \bauthor{\bsnm{Dennell},~\bfnm{R.}\binits{R.}}
(\byear{1998}).
\btitle{Harvesting the Sea, Farming the Forest: The Emergence of Neolithic
  Societies in the Baltic Region}.
\bpublisher{Sheffield Academic Press}, \blocation{Sheffield}.
\bptok{imsref}%
\end{bbook}
\endbibitem

\end{thebibliography}
\end{document}